\documentclass[prb,twocolumn,showpacs,floatfix,amsmath,amssymb]{revtex4}

\pdfoutput=1
\usepackage{ifpdf}

\ifpdf
\usepackage[pdftex]{graphicx}
\else
\usepackage[dvips]{graphicx}

\fi

\usepackage{latexsym,hyperref}
\usepackage{graphicx}
\usepackage{times}
\usepackage{amsmath}
\usepackage{dcolumn}
\usepackage{subfigure}
\usepackage{latexsym,amsmath,amssymb,bm}
\usepackage{euscript}
\def\etal{~\textit{et~al.}} 
\newcommand{\ket}[1]{| #1 \rangle}
\newcommand{\bra}[1]{\langle #1 |}

\newcommand{\be}{\begin{equation}}
\newcommand{\ee}{\end{equation}}

\begin{document}

\title{
Near zero modes in condensate phases of the Dirac theory on the honeycomb lattice.} 
\author{Doron L. Bergman and Karyn Le Hur}
\affiliation{Physics Department, Yale University, New Haven, CT 06520-8120}

\date{\today} 
 
\begin{abstract} 
We investigate a number of fermionic condensate phases on the
honeycomb lattice, to determine whether topological defects 
(vortices and edges) in these phases can support bound states 
with zero energy. We argue that topological zero modes bound 
to vortices and at edges are not only connected, but should 
in fact be \emph{identified}. Recently, it has been shown 
that the simplest s-wave superconducting state for the Dirac fermion
approximation of the honeycomb lattice at precisely half filling, 
supports zero modes inside the cores of vortices (P. Ghaemi and  F. Wilczek, 2007). 
We find that within the continuum Dirac theory the zero modes are not unique neither to this phase, nor to 
half filling. In addition, we find the \emph{exact} wavefunctions for vortex bound zero modes, as well as 
the complete edge state spectrum of the phases we discuss. 
The zero modes in all the phases we examine have even-numbered degeneracy, and as such pairs of any Majorana
modes are simply equivalent to one ordinary fermion. As a result, 
contrary to bound state zero modes in $p_x+i p_y$ superconductors, vortices here do \emph{not}
exhibit non-Abelian exchange statistics. The zero modes in the pure Dirac theory are seemingly topologically 
protected by the effective low energy symmetry of the theory, yet on the original 
honeycomb lattice model these zero modes are split, by explicit breaking of the effective low energy symmetry. 
\end{abstract} 

\maketitle 
 

\section{Introduction}
\label{intro}

In recent years, $p_x + i p_y$ fermionic condensate states, have received much attention
due to the expectation that vortices in this state will exhibit non-Abelian (Braiding) statistics
\cite{Read:prb00, Ivanov:prl01, Gurarie:prb07, Tewari:prl07, Tewari:prl07A,Read:2008},
and their potential applicability to topological quantum computing\cite{DasSarma:RMP07,Stern:2007}.
A frantic experimental effort to observe these effects is currently under 
way\cite{Dolev:2008,Radu:2008}.
The non-Abelian effects are caused by the presence of quasiparticle zero modes (states with energy precisely at 
the Fermi level) bound to vortex cores\cite{Kopnin:prb91, Volovik:93, Volovik:99, Volovik:book, Read:prb00, Ivanov:prl01, Gurarie:prb07, Tewari:prl07,Tewari:prl07A}. These zero modes appear at sample edges as well
\cite{Read:prb00,Stone:prb04,Fendley:prb07},
and we will refer to them collectively as topological zero modes.

Zero modes in any BCS mean field Hamiltonian\cite{Read:prb00, Ivanov:prl01, Gurarie:prb07}
can always be expressed as Majorana fermions. 
Pairs of Majorana states will combine to form single fermionic degrees of freedom,
which can then be occupied or not.
The $p_x + i p_y$ superconducting states allow
\emph{single} zero modes bound to vortices (of unit vorticity).
Therefore, fermionic states can only be formed by a superposition of two zero modes, bound to 
\emph{different} vortices. In this way, these fermionic modes provide a natural entangled state
between two spatially separated objects (the vortices)\cite{Ivanov:prl01}.
This entanglement is the source of 
the non-Abelian mutual statistics, when moving one vortex adiabatically around another.

The apparent rarity of $p_x + i p_y$ superconducting states 
has made difficult the effort to observe these zero modes in experiment.
It would therefore prove useful to have further candidate states for displaying 
non-Abelian statistics, which could then be searched for experimentally.

The topological zero modes in the $p_x + i p_y$ state, are found as solutions  
of a set of coupled Dirac-like Bogoliubov-de-Gennes (BdG) 
equations. The source of the Dirac-like behavior is the symmetry of the $p_x + i p_y$ 
superconducting order parameter.
An alternative way to end up with BdG equations of the form of a Dirac equation, 
one which does \emph{not} require the pairing function to be of the $p_x + i p_y$ form,
is to have a kinetic energy term that is of the Dirac form. 
The most celebrated example where this occurs is in the honeycomb lattice
tight-binding model, where close to half filling the band structure has 
Dirac-like dispersion, and the behavior of the system can be approximated
by two flavors of Dirac fermions. The effective Dirac like dispersion has been experimentally
observed in monolayer graphene\cite{geim2007rg}.
With the Dirac-like behavior of the BdG equations already guaranteed in this approximation, 
we are now free to ask whether zero modes exist in vortex cores of a whole variety of superconducting states
on the honeycomb lattice. The simplest state one could consider is the s-wave spin-singlet pairing
state. Some time ago\cite{Cugliandolo:1989}, it was shown that this same superconducting state in 
a (square) lattice model with Dirac dispersion near the Fermi energy supported zero modes bound 
to vortex cores. More recently\cite{Ghaemi:2007} zero modes were shown to exist (bound to vortex 
cores) in this state, in the Dirac continuum theory of the honeycomb lattice at precisely half filling. 
These findings are in stark contrast to the behavior of two dimensional s-wave superconductor vortices in 
fermionic systems with simple quadratic dispersion, where no zero modes exist\cite{Tewari:prl07,DeGennes:64}. 

Following this radically different result, in this article, we will investigate
other geometries and phases for possible presence of topological zero modes.
In addition, we will determine whether the zero modes appear in the actual
lattice model.

Even the simplest effective attraction between fermions can cause a superconducting state
to appear. However, the precise nature of the phase, namely the symmetry of the order parameter, depends on the 
details of the effective interaction. In Ref.~\onlinecite{Uchoa:prl07} it was shown in a simple mean field analysis that fermions on the honeycomb lattice paired in spin singlets may support not only an s-wave state, but also an
effective $p_x+i p_y$ state, as well as a mixed s-wave/$p_x+i p_y$ phase (and earlier work\cite{PhysRevB.63.134421} 
also suggested a p-wave superconducting state may appear in graphene).
Given the evidence of zero modes in the s-wave phase at half filling\cite{Ghaemi:2007}, it is interesting 
to explore whether the $p_x+i p_y$ spin-singlet phase may also support zero modes, as well as whether these 
zero modes are peculiar
to the half filling point (recently\cite{zhao:230404} it was shown that s-wave superconductivity is far
more likely to appear in a fermionic system in the honeycomb lattice away from half filling).
We will show in this manuscript that zero modes appear in the Dirac continuum theory in
both the s-wave and the $p_x+i p_y$ phase, even when deviating from the special half filling point 
(The $p_x+i p_y$ state is in fact gapped only when the fermions are \emph{away} from half filling).
We also demonstrate that the zero modes we find are 4-fold degenerate, 
and so fermionic modes can be formed by pairs on the \emph{same} vortex. The mechanism for 
entanglement between vortices is therefore unfortunately lost, and no non-Abelian effects 
are expected in these systems. Note that it is not obvious that no condensate phase of 
fermions on the honeycomb lattice will exhibit non-Abelian statistics.



As mentioned above, in the $p_x+i p_y$ state (with the regular quadratic kinetic energy)
it is known that the zero modes bound to vortex cores appear in conjunction with 
edge states zero modes. We argue in this manuscript that this connection is in fact quite general,
and that vortex core topological zero modes and edge state zero modes should in fact be \emph{identified}.
We demonstrate this in the superconducting states we analyze here, by finding the low energy edge states
of each phase. As expected, we find precise correspondence with the vortex core bound states - in both 
spin-singlet phases, there exist 4 zero modes.
A further signature of the identification of the vortex and edge states is that the wavefunctions
have all the same physical parameters - the same exponential decay lengths, as well the 
same oscillation length scales (when those exist).

A perhaps simpler indication of whether zero modes can appear at the edges or vortex cores
of a given superconducting (SC) state, is to consider the 
SNS (superconducting-normal-superconducting) junction\cite{Ashvin:2001},
with some phase difference $\phi$ between the SC droplets.
We find that as in the regular $p_x + i p_y$ state\cite{Ashvin:2001}, zero modes 
appear only when $\phi = \pi$.
As we will see, the edge state calculation we employ in the continuum limit
is limited in the type of honeycomb lattice edge it can be used for.
For this reason the SNS junction calculation is useful - 
it shows that at least within the continuum limit the precise alignment of the edge
is immaterial. We will find that the number of zero modes found in the SNS geometry 
is $8$ rather than $4$, giving us the first hint that details of boundary conditions
are important in this problem - the SNS junction geometry has extra symmetries,
compared with the edge and vortex cases.

The zero modes we will uncover in what follows, all have an even degeneracy. 
In general (for unit vorticity), only single SC quasiparticle zero modes are topologically protected to all 
possible perturbations, however, if there is a symmetry mandated degeneracy, which
is not broken by any of the perturbations, a degenerate set of zero modes can still 
be protected to perturbations, and hence topologically protected (modulo symmetry
mandated degeneracy). We begin the main body of our manuscript with a discussion of 
this distinction in general settings. 

In our case, the 4-fold degeneracy is mandated by the symmetries of the Dirac continuum 
theory. However, the full symmetry is \emph{not} present in the underlying lattice model,
and so there is a danger that the zero modes can appear in the continuum model, but 
\emph{not} in the lattice model. 
%
%
By numerical diagonalization of the lattice model, 
we find indeed this is the case - the zero modes do not exist in the lattice model.


The remainder of our manuscript is organized as follows.
In section~\ref{GenZeroModes} we discuss topological protection of zero modes
in conjunction with symmetry mandated degeneracy.
In section~\ref{vortex_edge_equiv} we present our general argument for identifying
zero modes bound to vortex cores and at sample edges.
We then proceed to section~\ref{Condensate_phases}, where we present the honeycomb
lattice BCS model, describe a number of possible superconducting states, 
and then set up a continuum limit for the model, which allows us to perform explicit calculations
looking for zero modes in the SNS junction (section~\ref{SNS}), the edge states (section~\ref{Edge_states}),
and finally in the vortex cores (section~\ref{Vortex_states}).
We present the results of a numerical calculation on the precise honeycomb lattice BCS model 
in section~\ref{Numerics}. In section~\ref{ExperimentalRealizations}
we proceed to discuss a possible experimental realization of superconducting states on the honeycomb lattice.
We then discuss Zeeman splitting in section~\ref{Magnetic_Field_Splitting},
and propose how it can be used to test the physics we describe here
experimentally, using the absorption spectrum of the system.
We conclude our manuscript with the discussion in section~\ref{discussion}.

\section{Zero modes in BCS Hamiltonians modulo symmetry mandated degeneracies}
\label{GenZeroModes}

The most general BCS\cite{BCS:1957} Hamiltonian has the form
\be\label{genBCS}
{\mathcal H}_{\textrm{BCS}} = \sum_{a b} 
\left[
f_a^{\dagger} h_{a b} f_b^{\phantom\dagger} -
\frac{1}{2}
f_a^{\phantom\dagger} \Delta_{a b} f_b^{\phantom\dagger} - \frac{1}{2}
f_b^{\dagger} \Delta_{a b}^* f_a^{\dagger}
\right]
\; ,
\ee
where $a,b$ are generalized coordinates, and may include spin, position and any other 
degree of freedom one can imagine. The fermionic operators $f_a$ satisfy 
standard anticommutation relations.
The operator $h$ is hermitian, and the pairing function must be anti-symmetric in the generalized coordinates 
$\Delta_{a b} = -\Delta_{b a}$. A Bogoliubov transformation 
$\gamma_E^{\dagger} = \sum_a \left[ u_{E a} f_a^{\dagger} + v_{E a} f_a^{\phantom\dagger} \right]$
diagonalizes this quadratic Hamiltonian. From the eigenstate equations 
$\left[ {\mathcal H}_{\textrm{BCS}} , \gamma_{E}^{\dagger}\right] = E \gamma_{E}^{\dagger}$
the Bogoliubov-de-Gennes (BdG) equations are derived
\be\label{genBdG}
\left(
\begin{array}{cc}
h & \Delta \\
\Delta^{\dagger} & -h^{T}
\end{array}
\right) \cdot \psi
= E \psi
\; ,
\ee
where $\psi = (u_a, v_a)$ (we drop the $E$ index to avoid clutter).
The BdG equations are then simply the eigenvalue problem
${\mathcal H}_{\textrm{BdG}} \psi = E \psi$, in terms of the BdG Hamiltonian
${\mathcal H}_{\textrm{BdG}}$. This operator has the symmetry
$ \omega^x {\mathcal H}_{\textrm{BdG}} \omega^x  = - {\mathcal H}_{\textrm{BdG}}^*$, where $\omega^x$
is the x-Pauli matrix in the so-called Nambu spinor basis $\omega^{x,y,z}$ (see table~\ref{tab:PauliMatrixSets}),
acting on the $(u,v)$ components of ${\mathcal H}_{\textrm{BdG}}$.
The Nambu spinor obeying $\omega^z \psi= + \psi$
corresponds to a pure fermion ($v_a = 0$), while $\omega^z \psi= - \psi$
corresponds to a pure hole ($u_a = 0$).
This relation tells us that given an eigenvector $\psi$ with eigenvalue $E$,
$\omega^x \psi^*$ will also be an eigenvector with the eigenvalue $-E$.
The matrix eigenvalue equation \eqref{genBdG} can indeed yield both positive and 
negative eigenvalues, however it is important to note that in this formulation of the BdG
equations, the operators $\gamma_E^{\dagger} = \gamma_{-E}^{\phantom\dagger}$,
and so these pairs of $\pm E$ energy states are \emph{not} independent\cite{Ivanov:prl01}.
This is a consequence of the doubling of the number of degrees of freedom in \eqref{genBCS}\cite{Altland:prb96}.

\begin{table}
	\centering
		\begin{tabular}{| l || c | c | c | c |}
		\hline
    Spinor type & Nambu  & Spin   & Sublattice & Dirac valley \\
       & spinor & spinor & spinor     & spinor \\ \hline
    Pauli matrices & $\omega^{x,y,z}$ & $\sigma^{x,y,z}$ & $\eta^{x,y,z}$ & $\tau^{x,y,z}$ \\
    index variables & - & $\alpha,\beta$ & $\mu,\nu$ & $A,B$ \\
    index values & - & $\uparrow,\downarrow$ & $1,2$ & $R,L$ \\			
		\hline
		\end{tabular}
	\caption{Pauli matrix sets}
	\label{tab:PauliMatrixSets}
\end{table}

The BCS ground state is a state that is annihilated by $\gamma_E \ket{BCS}$ for $E>0$,
and by $\gamma_E^{\dagger}\ket{BCS}$ for $E<0$, which really are the same set of operators.
Naively, one would be tempted to think of the ground state as a fermi sea of quasiparticle levels, 
all occupied below the Fermi level $E<0$, and all unoccupied for levels above $E>0$.
However, as we see here, the ``hole'' $\gamma_{-E}^{\phantom\dagger} \ket{BCS}$ 
and particle $\gamma_E^{\dagger} \ket{BCS}$ excitations (here $E>0$) are in fact the \emph{same},
and so the the BCS ground state should \emph{not} be thought of as a filled Fermi sea of Bogoliubov 
quasiparticles. 
In the physical interpretation of the spectrum of the BdG equations, only
the eigenvectors with energies $E\geq 0$ should be understood as wavefunctions of physical excitations.

Despite the subtleties of the physical interpretation of the BdG equations in this form, 
the purely mathematical analysis of them as an eigenvalue problem, is extremely useful in identifying
topologically protected zero modes. The BdG Hamiltonian (the matrix-operator of \eqref{genBdG})
has a spectrum with $\pm E$ energy pairs. For zero modes ($E=0$), if some perturbation to the BdG 
matrix-operator were to cause the zero mode eigenstate to acquire nonzero energy, there would have to be 
another zero mode eigenstate acquiring the \emph{opposite} nonzero energy. 
The Atiyah-Singer index theorem\cite{Atiyah:63} tells us that the number of solutions to an 
eigenvalue problem cannot be changed when deforming the operator continuously.
Plainly put, new eigenstates cannot appear out of thin air - they must be deformations of eigenstates of the unperturbed system. Therefore, if the zero modes appear in pairs, they can in principle split. 
However, if the number of zero modes is \emph{odd}, then at least one zero mode cannot be split by any
physical perturbation (those that preserve the general form of the BCS Hamiltonian \eqref{genBCS}), and 
it is thus topologically robust.

The most celebrated example of this, is the $p_x +i p_y$ state, in which there is \emph{one} zero mode bound to
a (unit vorticity) vortex\cite{Read:prb00} core, and so it is topologically protected. If the BdG matrix-operator
spectrum has some degeneracy mandated by symmetries of the system, that are conserved by physical perturbations,
then the same degeneracy must hold. It then suffices for the zero modes to have an odd number modulo the minimal degeneracy mandated by the symmetries of the system, in order to be topologically protected - deviation 
from zero energy would split the zero modes in half, and the degeneracy protected by the symmetry would be 
violated. It is however important to realize that if the deformation breaks the
symmetries dictating the degeneracy of the entire spectrum, the zero modes may split. 
The zero modes are therefore protected by the \emph{combination} of symmetry and topological protection.

\section{Equivalence of vortex bound states and edge states}
\label{vortex_edge_equiv}

In this section we will argue that quite generally the vortex core bound zero mode states should be 
\emph{identified} with zero mode edge states, in condensate phases. The connection between the 
presence of zero modes at vortex cores and at sample edges has been previously examined in the 
context of $p_x + i p_y$ condensates of quadratic-dispersion fermions\cite{Read:prb00,Fendley:prb07}.

\begin{figure}
	\centering
		\includegraphics[width=2.0in]{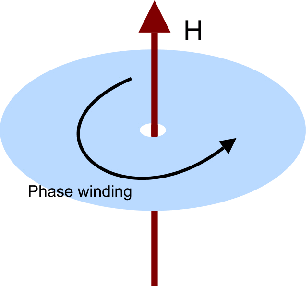}
	\caption{The punctured infinite plane geometry, with a magnetic flux threaded through the puncture (a vortex).}
	\label{fig:Vortex_puncture}
\end{figure}

Consider the infinite plane with the order parameter amplitude non-zero and radially uniform only in 
the range $r > L$. Note that in the case of the $p_x +i p_y$ condensate the order parameter
has the form\cite{Read:prb00} $\{ \Delta, \partial_x + i \partial_y \}$, and we refer to
$\Delta$ as the condensate amplitude.
The order parameter will include a phase winding $\Delta = |\Delta| e^{i m \phi}$ (here 
and throughout this section, $m$ is an integer), 
so that the region $r < L$ at the disc center models a vortex core. 
A state bound to the vortex core will have a radial profile for the wavefunction that is 
decaying exponentially $\sim e^{- \lambda r}$ in the superconducting region $r> L$. 
In order for the wavefunction to be normalizable in the $r \rightarrow 0$ limit as well, 
the wavefunction amplitude must have the form $|\psi| \sim r^{\nu}$ with $\nu > -1$, in this limit.
For $\nu = -1$, the wavefunction norm $\int_0^a |\psi|^2 r dr \sim \log{\frac{a}{\epsilon}}|_{\epsilon \rightarrow 0} $
diverges logarithmically. We have therefore uncovered another boundary condition on the 
wavefunction - $\psi r$ must vanish at $r=0$.
Since bound state zero modes should not be sensitive to the detailed boundary condition inside the 
vortex core\cite{Read:prb00}, 
the $r=0$ point can then be mapped onto a hard wall of a small radius $r = \delta$,
on which $\psi r = 0$, and therefore $\psi = 0$ on this wall. An even simpler model of the vortex
is obtained when we identify $\delta = L$. The geometry then consists of a punctured 
infinite disc (see Fig.~\ref{fig:Vortex_puncture}), with the condensate order parameter amplitude
having uniform magnitude, and including a phase winding.

\begin{figure}
	\centering
		\includegraphics[width=1.5in]{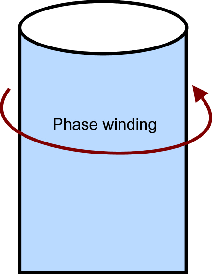}
	\caption{The semi-infinite cylinder geometry, topologically equivalent to the punctured infinite plane.
	There is a phase winding around the cylinder, corresponding to the phase winding of the vortex.}
	\label{fig:cylinder}
\end{figure}

\begin{figure}
	\centering
		\includegraphics[width=1.7in]{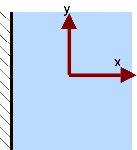}
	\caption{The semi-infinite plane geometry, topologically equivalent to the punctured infinite plane.}
	\label{fig:semi_infty}
\end{figure}

The geometry of the punctured infinite plane can be continuously deformed into a semi-infinite
cylinder geometry (see Fig.\ref{fig:cylinder}), with radius $L$. Taking $L \rightarrow \infty$ then
turns the geometry into the semi-infinite plane, Fig.\ref{fig:semi_infty}. The boundary condition 
at $r = L$ corresponds now to a hard wall sample edge $\psi|_{\textrm{wall}} = 0$.
The radial exponentially decaying solution of the punctured infinite plane geometry will deform 
into an exponentially decaying solution in the direction perpendicular to the edge (see 
Fig~\ref{fig:semi_infty}), and the decay length will remain the \emph{same} in both geometries
(in terms of the physical length scales in the system).

The phase winding can be ignored in the formal limit $L \rightarrow \infty$,
but must be taken into account when considering a finite size of the plane.
The smooth deformation we employed to map between the vortex and an edge,
will also map $\Delta = |\Delta| e^{i m \phi} \rightarrow
\Delta = |\Delta| e^{i m 2 \pi \frac{y}{L}} $. The phase winding in the order parameter will 
change the bound state wavefunctions qualitatively. 
When going from $y$ to $y+L$ we will pickup
the requisite phase of $2 \pi m$.


With the insight from the last section, it is now clear that modulo symmetries that
are not related to position space the mapping we have described here, \emph{identifies} the 
vortex core topological zero modes and topological zero mode edge states.
Furthermore, we may conclude that showing the existence of one implies the existence of the other.
Indeed, for the $p_x+i p_y$ superconducting phase it is known\cite{Read:prb00, Stone:prb04, 
Fendley:prb07} that zero mode edge states exist, in the presence of a vortex.

It is instructive to examine how the momentum quantization evolves when mapping between the vortex and the 
edge geometries. Consider the semiclassical limit introduced in Ref.~\onlinecite{Ashvin:2001}, where the quantization of angular/linear momenta can be inferred from a Bohr-Sommerfeld quantization rule of the classical orbits
\be\label{Bohr_Sommerfeld1}
2 \pi \ell \hbar = 
\oint {\bf p} \cdot d{\bf x}
+ \frac{\hbar}{2} \oint {\mathcal A}(\hat n) \cdot d{\hat n}
\; .
\ee
Where here and in what follows $\ell$ is integer,  and
the second integral on the RHS is the Berry phase traced by the 
(classical) Nambu vector ${\hat n}$. Note that as opposed to Ref.~\onlinecite{Ashvin:2001},
we do not transform to the London gauge.
This is the result for a single band of quadratically-dispersing fermions,
and we assume that a similar quantization rule will appear when taking a similar
semi-classical limit of the more general problem. In particular
\be\label{Bohr_Sommerfeld2}
2 \pi \ell \hbar = 
\oint {\bf p} \cdot d{\bf x}
+ \textrm{topologically invariant terms ...}
\; .
\ee
Then in the process of deforming from the vortex to the edge, only the 
first term on the RHS changes. For a rotationally symmetric system, if the the angular 
momentum of the vortex core bound states is quantized $R p = \hbar (\ell + \gamma)$ 
($0 \leq \gamma < 1$), then the linear momentum for the edge states 
will be quantized as $q = \frac{2 \pi  \hbar}{L} (\ell + \gamma)$ ( where $L = 2\pi R$ is as before, the 
system size in the direction parallel to the edge).

Typically, the edge state energies $E \sim q$ at low momenta/energy, and so we will 
need $\gamma = 0$ or integer angular 
momentum states in the vortex core, in order to support zero modes.

\section{Condensate phases}
\label{Condensate_phases}

In this section we briefly present the variety of condensate phases we will be examining in this manuscript.

We begin by considering a simple model of fermions on the honeycomb lattice,
either spinless or including spin. We include nearest neighbor hopping (of strength $t$) 
and fermion density-density interactions
\be
\begin{split}
{\mathcal H} = & -t \sum_{\langle i j \rangle \alpha} f_{i \alpha}^{\dagger} f_{j \alpha}^{\phantom \dagger} +
\mu \sum_{j \alpha} f_{j \alpha}^{\dagger} f_{j \alpha}^{\phantom \dagger}
\\ 
+ &
\sum_{i j, \alpha \beta} f_{i \alpha}^{\dagger} f_{i \alpha}^{\phantom \dagger} V_{i j}^{\alpha \beta} 
f_{j \beta}^{\dagger} f_{j \beta}^{\phantom \dagger}
\; ,
\end{split}
\ee
where $f_{j \alpha}$ are the bare fermionic operators.
The indices $i,j$ run over the sites of the honeycomb lattice, and the greek 
letters $\alpha, \beta = \uparrow, \downarrow$ denote the spin indices.
We point out that we neglect the gauge field in all our calculations.
For spinless fermions, the indices $\alpha,\beta$ should be dropped.
The interaction matrix is symmetric $V_{i j}^{\alpha \beta} = V_{j i}^{\beta \alpha}$,
and is chosen such that $V_{i i}^{\alpha \alpha} = 0$, so that $\mu$ indeed will be the Fermi energy.

Throughout this manuscript we will assume we are in the weak interaction limit $t \gg V_{i j}$,
so that BCS mean field theory is applicable.

\subsection{Order parameters}

\begin{figure}
	\centering
		\includegraphics[width=2.0in]{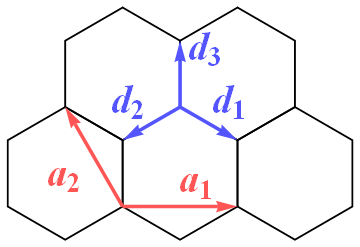}
	\caption{Conventions for the honeycomb lattice. The (blue) vectors marked $d_{1,2,3}$ represent the nearest neighbor vectors from sublattice 1 to sublattice 2, and the (red) vectors $a_{1,2}$ mark the Bravais lattice vectors.
	In our conventions, the nearest neighbor distance is set to 1.
}
	\label{fig:HoneyComb_Conventions_pic}
\end{figure}

An order parameter for off-diagonal long range order can be chosen as
$\Delta_{i j}^{\alpha \beta} = - 2 V_{i j}^{\alpha \beta} \langle \left( f_{i \alpha} f_{j \beta} 
\right)^{\dagger} \rangle$
and must be antisymmetric $\Delta_{i j}^{\alpha \beta} = - \Delta_{j i}^{\beta \alpha}$. 
For spinless fermions, or spin triplet ($\Delta$ symmetric in spin indices) condensates, 
the order parameter is anti-symmetric in the lattice sites, and \emph{must} break parity
(the order parameter becomes anti-symmetric in swapping the $i,j$ indices). 
A mean field BCS Hamiltonian can then be obtained
\be\label{lattice_model}
\begin{split}
{\mathcal H}_{BCS} = & -t \sum_{\langle i j \rangle \alpha} f_{i \alpha}^{\dagger} f_{j \alpha}^{\phantom \dagger} +
\mu \sum_{j \alpha} f_{j \alpha}^{\dagger} f_{j \alpha}^{\phantom \dagger}
\\ 
- &
\frac{1}{2}
\sum_{i j, \alpha \beta} \left[ f_{i \alpha}^{\phantom \dagger} f_{j \beta}^{\phantom \dagger} 
\Delta_{i j}^{\alpha \beta} + h.c. \right]
\; .
\end{split}
\ee
In momentum space the BCS mean-field Hamiltonian reads
\be\label{BCS1}
\begin{split}
{\mathcal H}_{\textrm{BCS}} =
\sum_{\bf q} \sum_{\mu \nu}
\Bigg[ & 
\sum_{\alpha}
f_{\mu \alpha}^{\dagger}({\bf q}) 
\left( \mu \delta_{\mu \nu}  - t \, \Gamma({\bf q})_{\mu \nu} \right)
f_{\nu \alpha}^{\phantom\dagger}({\bf q})
\\ - &
\frac{1}{2}
\sum_{\alpha \beta}
\left[ f_{\mu \alpha}^{\phantom \dagger}({\bf q}) 
\Delta_{\mu \nu}^{\alpha \beta}({\bf q}) f_{\nu \beta}^{\phantom \dagger}(-{\bf q}) 
+ h.c. \right]
\Bigg]
\; ,
\end{split}
\ee
where $f_{\mu \alpha}({\bf q})$ are the bare fermionic operators in momentum space,
and we have introduced the matrix 
\be\label{gamma}
\Gamma({\bf q}) = 
\left(
\begin{array}{cc}
0 & \gamma({\bf q}) \\
\gamma({\bf q})^* & 0 \\
\end{array}
\right)
\; .
\ee
The indices $\mu, \nu = 1,2$ denote the two triangular sublattices of the honeycomb lattice.
Note that while we use $\mu$ for both a sublattice index, and for the chemical potential,
it should be clear from context when $\mu$ is used for one or the other (specifically,
whenever both appear in the same equation, the index $\mu$ is always a subscript).
Furthermore, $\gamma({\bf q}) = \sum_{\ell=1}^3 e^{+i {\bf q} \cdot {\bf d}_{\ell} }$
where ${\bf d}_{1,2,3}$ are the three vectors from any given site in sublattice 1 
to it's 3 nearest neighbors on sublattice 2 (see Fig.~\ref{fig:HoneyComb_Conventions_pic} 
for an illustration of our conventions). 
Finally, momentum is summed over the first Brillouin zone.

For spinless fermions or spin-triplet condensates, the parity broken $\Delta_{i j}$ implies 
$\Delta_{\mu \nu}^{\alpha \beta}({\bf q}) = - \Delta_{\nu \mu}^{\alpha \beta}(-{\bf q})$.
We now turn to several condensate order parameters of interest.

First, we mention the spin-singlet condensate phases introduced in Ref.~\onlinecite{Uchoa:prl07},
which take the form $\Delta_{i j}^{\alpha \beta} = \Delta_{i j} i \sigma^y_{\alpha \beta}$.
Here and elsewhere we will use the notation $\sigma^{x,y,z}$ for the spin Pauli matrices (see table~\ref{tab:PauliMatrixSets}).
The function $\Delta_{i j}$ is then symmetric, and Uchoa et al.\cite{Uchoa:prl07} take
$\Delta_{i j} = \delta_{i j} \Delta_0$ for an s-wave order parameter, and 
$\Delta_{i j} = \Gamma_{i j} \frac{1}{2} \Delta_1$, where $\Gamma_{i j}$ is the adjacency matrix for 
the honeycomb lattice (takes a value of $1$ for nearest neighbor sites, and zero otherwise).
The $\Delta_1$ order parameter mimics the structure of the tight binding kinetic energy term
for the honeycomb lattice, $\sum_{\langle i j \rangle} \ldots = \frac{1}{2} \sum_{i j} \Gamma_{i j} \ldots$,
and in fact the Fourier transform of $\frac{1}{2} \Gamma_{i j}$ is simply the matrix $\Gamma({\bf q})$ of \eqref{gamma}.
As a result, near half filling just as the tight binding term has two Dirac nodes, so does the 
order parameter $\Delta_1$. The $\Delta_1$ order parameter then has the approximate form of 
a $p_x+i p_y$ order parameter\cite{Uchoa:prl07}. It may be a bit surprising to find a $p_x+i p_y$
pairing function in a spin-singlet condensate, since it is then anti-symmetric under both spin 
exchange, and momentum inversion. However, the additional structure from the sublattice basis 
provides a third antisymmetric component of the pairing function, that keeps the overall anti-symmetry. It is important to emphasize at this point that this $p_x+i p_y$ phase
is gapped only when we are away from half filling ($\mu \neq 0$).

Now we turn to spinless/spin-triplet order parameters. It is most convenient to write the 
spin-triplet order parameter in the form
$
\Delta_{i j}^{\alpha \beta} = \left[
i \sigma^y {\vec \sigma} \cdot {\vec d}_{i j}
\right]_{\alpha \beta}
$. Here the (3-component) vector is antisymmetric ${\vec d}_{i j} = -{\vec d}_{j i}$.
Let us focus on a single component of the vector ${\vec d}_{i j}$, to simplify our analysis, and also
because this is equivalent to the spinless fermion case. Let us denote this single component as
$\Delta_{i j}$, which is still anti-symmetric. The Fourier transform of this
pairing function is a matrix $\Delta_{\mu \nu}({\bf q})$. The sublattice structure now mimics the 
behavior of the spin matrix structure, and can allow both sublattice spinor-singlet as well as 
triplet structures. We introduce a new set of Pauli matrices $\eta^{x,y,z}$ in the 2-sublattice space 
(see table~\ref{tab:PauliMatrixSets}),
and now the spinless fermion pairing function can be written in complete generality as
$
\Delta_{\mu \nu}({\bf q}) = \left[
i \eta^y {\vec \eta} \cdot {\vec \Delta}({\bf q})
+
i \eta^y \Delta_0({\bf q})
\right]_{\mu \nu}
$. As mentioned earlier in this section, the order parameter for spinless
fermions will satisfy $\Delta_{\mu \nu}({\bf q}) = - \Delta_{\nu \mu}(-{\bf q})$,
implying that $\Delta_0({\bf q}) = \Delta_0(-{\bf q})$
and ${\vec \Delta}({\bf q}) = - {\vec \Delta}(-{\bf q})$.

The simplest momentum structure in the order parameter $\Delta_{\mu \nu}({\bf q})$ would be just a function uniform
in momentum space. A valid antisymmetric s-wave order parameter for spinless fermion pairing is
$\Delta_{\mu \nu}({\bf q}) = 
\left(
\begin{array}{ll}
0 & -i \Delta_0 \\
i \Delta_0 & 0
\end{array}
\right) = 
\eta^y \Delta_0
$.
However, attempting to transform this order parameter back to real space reveals that it break the 
honeycomb lattice symmetries. Namely, the pairing function $\Delta_{i j}$ is non-zero only for 
$i,j$ on different sublattices, and in the same unit cell. The other nearest neighbors pairs of either
$i,j$ do not enjoy a pairing amplitude, and so discrete rotation symmetry is broken.

For an order parameter that is linear in momentum (at least in a continuum limit),
and maintains all the symmetries of the honeycomb lattice ( apart from the inversion symmetry, 
as mentioned above), we take $\Delta_{i j}$ nonvanishing only on nearest-neighbor links.
We choose for the directed links from one sublattice to the other the value $\Delta_{i j} = \Delta_2$,
and opposite for $\Delta_{j i} = - \Delta_2$. In momentum space, the order parameter yields
\be
\Delta_{\mu \nu} =
\Delta_2 \left(
\begin{array}{cc}
0 & \gamma({\bf q}) \\
- \gamma({\bf q})^* & 0
\end{array}
\right)_{\mu \nu}
\; .
\ee
This pairing function can be conveniently rewritten as
$\Delta_{\mu \nu} = \Delta_2 \eta^z \Gamma({\bf q})$. 
We note that this pairing function 
includes parts that are symmetric as well as anti-symmetric in the sublattice space.
This pairing function is a directed version of the link-pairing 
order parameter in Ref.~\onlinecite{Uchoa:prl07}, and similarly, when the fermions in the 
system are near half filling, a $p_x + ip_y$ structure appears from the matrix $\Gamma({\bf q})$,
and the order parameter has the approximate symmetry of a $p_x + i p_y$ order parameter.

With the spinless $p_x+i p_y$ pairing,
the bulk energy spectrum for the precise lattice model is found to be
$
E = \pm
\left[
\left( t^2 + |\Delta_2|^2\right) |\gamma|^2
+ \mu^2
\pm t |\gamma| \sqrt{
|\Delta_2|^2 \left( \gamma + c.c \right)^2
+ 4 \mu^2
}
\right]^{1/2}
$
and it has nodes at the points
${\bf q} = p \left( 1,\frac{1}{\sqrt{3}} \right) $
where 
$p = \arccos{\left( \frac{|\mu|}{2 \sqrt{t^2 - |\Delta_2|^2}} - \frac{1}{2} \right)}$,
and symmetry related points. These points satify $\gamma$ \emph{real}, and 
$|\gamma|^2 = \frac{\mu^2}{t^2 - |\Delta|^2}$.
Regardless of where the Fermi surface is, the nodal points are near the Fermi surface
defined by $|\gamma|^2 = \frac{\mu^2}{t^2}$, since $|\Delta| \ll t$.
A superconducting phase with nodes very close to the Fermi surface allows for 
bulk states below the momentum space averaged gap. Our analysis in appendix~\ref{app:spinless}
will indeed confirm this.

In the following sections we will perform an exhaustive analysis of the 2 spin-singlet pairing phases to determine
whether they allow topological zero modes. The analysis of the spinless phase we leave to appendix~\ref{app:spinless} becuase this phase has gapless bulk excitations.

\subsection{Continuum limit}
\label{ContinuumLimit}

Analyzing the bound states in a vortex or an edge, is most easily done in a continuum limit of the lattice model.
Here we will follow the conventions of Ref.~\onlinecite{Alicea:prb06}, and work in the so-called 
``valley-isotropic'' convention of the honeycomb lattice fermionic models near half filling. 
We demonstrate how this continuum limit is used on the BCS 
Hamiltonian\cite{BCS:1957} for the various condensate states we will consider here.

\begin{figure}
	\centering
		\includegraphics[width=2.0in]{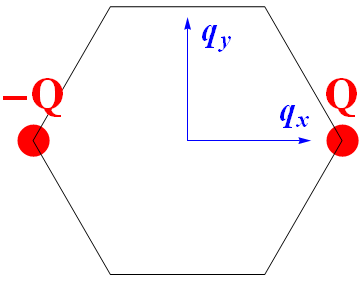}
	\caption{First Brillouin zone of the honeycomb lattice, showing the two Dirac node positions $\pm {\bf Q}$ (in red),
	and the main (x,y) axes indicated (in blue).}
	\label{fig:BZ}
\end{figure}

The Dirac nodes are a pair of points where $\gamma(\pm {\bf Q}) = 0$ (see Fig.~\ref{fig:BZ} for illustration).
In our conventions, the Dirac nodes are positioned at 
${\bf Q} = \left(\frac{4 \pi}{3 \sqrt{3}} ,0 \right)$ and $- {\bf Q}$.
The fermi operators are expanded about these two nodes\cite{CastroNeto:RMP07}, 
in the so-called ``valley-isotropic'' convention, 
and the two modes are identified as right(R) and left(L)
\be\label{continuum}
\begin{split} &
f_{\mu \alpha}({\bf q + Q}) \approx \Pi \psi_{\mu \alpha R}({\bf q})
\\ &
f_{\mu \alpha}({\bf q - Q}) \approx \Pi i \eta^y_{\mu \nu} \psi_{\nu \alpha L}({\bf q})
\end{split}
\; .
\ee
Here $\Pi$ is some normalization, 
and we have used the y-Pauli matrix $\eta^y$ acting in the 
sublattice ($\mu, \nu$) spinor space, that was introduced in the previous section.

In addition to the Nambu, spin and sublattice spinor Pauli matrices, we now introduce a fourth
set of Pauli matrices $\tau^{x,y,z}$ that act in the Dirac valley ($A,B = R,L$) spinor space
(see table~\ref{tab:PauliMatrixSets}). 
For clarity, from this point on, whenever it is convenient we will suppress indices which are being summed over. 

We expand the kinetic energy about the 2 Dirac nodes, using
$\gamma({\bf q} \pm {\bf Q}) \approx \mp \frac{3}{2} \left( p_x \mp i p_y \right) 
+ \frac{3}{8} \left( p_x \pm i p_y \right)^2 + \ldots$, to obtain
\be\label{Gamma_approx}
\Gamma({\bf q} \pm {\bf Q}) \approx \mp \frac{3}{2} \left( \eta^x q_x \pm \eta^y q_y \right)
+ \frac{3}{8} \left[ \eta^x \left(q_x^2 - q_y^2 \right) \mp 2 \eta^y q_x q_y \right]
\; .
\ee
Organizing the expressions using the various Pauli matrix sets we have introduced,
and using the continuum Fourier transform
\be\label{Fourier}
\psi_{\mu \alpha A}({\bf q}) = \int d^2 {\bf x} e^{-i {\bf q} \cdot {\bf x} } \psi_{\mu \alpha A}({\bf x})
\; ,
\ee
we derive a real space continuum version of the kinetic energy
\be\label{kinetic}
\begin{split}
{\mathcal H}_{\textrm{kinetic}} =
\int d^2 {\bf x} \, \psi^{\dagger}
\Big[ & \mu 
- i v \left({\vec \eta} \cdot \nabla \right) 
\\ + &
\frac{v}{4} \tau^z \left( \eta^x \left( \partial_x^2 - \partial_y^2 \right) - 2 \eta^y \partial_x \partial_y \right)
\Big] \psi 
\; ,
\end{split}
\ee
where ${\vec \eta} = {\hat x} \, \eta^x + {\hat y} \, \eta^y$, and $v = t \frac{3}{2}$.

At this point we observe that keeping only the linear derivatives, we obtain the celebrated
Dirac operator, and at that level of approximation the Kinetic term has an $SU(2)$
symmetry of valley spinor ($\tau^{x,y,z}$) rotations, in addition to the spin $SU(2)$ symmetry.  
The last term is a quadratic correction to the Dirac operator, that explicitly \emph{breaks}
the Dirac spinor $SU(2)$ invariance, reducing it to a $U(1)$ symmetry of rotations about $\tau^z$. 
This correction, while often ignored, will be examined in the calculations we perform here.

Next we turn to the pairing term in the BCS Hamiltonian.
With the three pairing phases we outlined in the previous section, and using the various
Pauli matrices we defined to compactify the expressions, the lattice Hamiltonian pairing term is
\be\label{pairing}
\begin{split}
{\mathcal H}_{\textrm{pairing}} = -\frac{1}{2}
\sum_{\bf q}
\Big\{
f({\bf q}) 
\Big[ &
i \sigma^y \left( 
\Delta_0 + \Delta_1 \Gamma({\bf q})
\right) 
\\ + & 
\Delta_2 \eta^z \Gamma({\bf q})
\Big]
f(-{\bf q}) 
+ h.c. \Big\}
\; .
\end{split}
\ee

In the continuum limit, we approximate BCS off-diagonal terms in the following manner
\be
\begin{split} &
\sum_{\bf p} f({\bf p}) M({\bf p}) f(-{\bf p})
\\ \approx &
\sum_{\bf q} \Big[ f({\bf q+Q}) M({\bf q + Q}) f(-{\bf q} - {\bf Q})
\\ & 
+ f({\bf q-Q}) M({\bf q-Q}) f(- {\bf q} + {\bf Q})
\Big]
\\ \approx & \Pi^2
\sum_{\bf q} \Big[ \psi_R({\bf q}) M({\bf q + Q}) i \eta^y \psi_L(-{\bf q})
\\ &
+ \psi_L({\bf q}) (-i) \eta^y M({\bf q-Q}) \psi_R(- {\bf q})
\Big]
\; ,
\end{split}
\ee
where $M({\bf p})$ is the pairing function.
In our case, the pairing function from \eqref{pairing} is most conveniently cast as 
\be
M({\bf p}) = 
i \sigma^y \left( \Delta_0 + \Delta_1 \Gamma \right) + \Delta_2 \eta^z \Gamma
\; .
\ee
Using the lowest order in the expansion \eqref{Gamma_approx}, 
the identity 
$ \eta^y \left( {\vec \eta}^* \cdot {\bf q} \right) \eta^y = -  \left( {\vec \eta} \cdot {\bf q} \right)$,
and absorbing a factor
of $\frac{3}{2}$ into both $\Delta_{1,2}$, we find
\be
\begin{split} &
M({\bf q + Q}) i \eta^y \approx
\left[
i \sigma^y \left( \Delta_0 - \Delta_1 ({\vec \eta} \cdot {\bf q}) \right) - \Delta_2 \eta^z {\vec \eta} \cdot {\bf q}
\right] i \eta^y
\\ &
-i \eta^y M({\bf q - Q}) \approx
\left[
i \sigma^y \left( - \Delta_0 + \Delta_1 ({\vec \eta} \cdot {\bf q}) \right) - \Delta_2 \eta^z {\vec \eta} \cdot {\bf q}
\right] i \eta^y
\; .
\end{split}
\ee
Combining these results, we find the pairing term in the Hamiltonian becomes in the continuum limit
\be
\begin{split}
{\mathcal H}_{\textrm{pairing}} = -\frac{1}{2}
\sum_{\bf q}
\Big\{
\psi({\bf q}) 
\Big[
& i \sigma^y \tau^y \eta^y \Delta_0 
-  \sigma^y \tau^y \Delta_1 ({\vec \eta} \cdot {\bf q}) \eta^y
\\ 
+ & \tau^x \Delta_2 \eta^z ({\vec \eta} \cdot {\bf q}) \eta^y 
\Big]
\psi(-{\bf q}) 
+ h.c. \Big\}
\; .
\end{split}
\ee

In order to be able to take slowly spatially-varying order parameter amplitudes ($\Delta_{0,1,2}$),
we need to reorganize the pairing term in real space as
\be\label{final_pairing}
\begin{split}
{\mathcal H}_{\textrm{pairing}} = - \frac{1}{2}
\int_{\bf x}
\Big\{
\psi
\Big[
 & i \sigma^y \tau^y \eta^y \Delta_0 
-   i \sigma^y \tau^y \frac{1}{2} \{ \Delta_1 , ({\vec \eta} \cdot {\nabla}) \eta^y \}
\\ 
+ & i \tau^x \frac{1}{2} \{ \Delta_2 , \eta^z ({\vec \eta} \cdot {\nabla}) \eta^y \}
\Big]
\psi
+ h.c. \Big\}
\; ,
\end{split}
\ee
where all the operators $\psi$ are function of ${\bf x}$, and $\{ .. , .. \}$ denotes the anti-commutator.

\subsection{Bogoliubov-de-Gennes equations}

With the final continuum forms of the kinetic \eqref{kinetic} and pairing \eqref{final_pairing} terms of the
BCS Hamiltonian, we can derive the BdG equations following the details of section~\ref{GenZeroModes}.
The BdG equations for the phases we examine in this manuscript take the form
\be
{\mathcal H}_{\textrm{BdG}} \cdot
\left(
\begin{array}{c}
u \\ v	
\end{array}
\right) = E 
\left(
\begin{array}{c}
u \\ v	
\end{array}
\right)
\; ,
\ee
where 
$
{\mathcal H}_{\textrm{BdG}} = {\mathcal H}_0 + {\mathcal H}_1 + {\mathcal H}_2 + {\mathcal H}_3
+ {\mathcal H}_4
$.
Here the kinetic term is
\be\label{H_0}
\begin{split} 
{\mathcal H}_0 & = 
\left(
\begin{array}{cc}
\left[ \mu -i v \left( {\vec \eta} \cdot \nabla  \right) \right] & 0 \\
0 & - \left[ \mu +i v \left({\vec \eta}^* \cdot \nabla \right) \right]
\end{array}
\right) 
\\ & = 
\mu \omega^z - i v \left( \eta^x \partial_x +\eta^y \omega^z \partial_y \right)
\\ & = 
\mu \omega^z - i v \hat{D}
\; ,
\end{split}
\ee
where we have introduced $\hat{D} = \left( \eta^x \partial_x +\eta^y \omega^z \partial_y \right)$.
The singlet s-wave pairing term is
\be\label{H_1}
\begin{split} 
{\mathcal H}_1 & = 
\left(
\begin{array}{cc}
0 & +i \Delta_0 \eta^y \sigma^y \tau^y \\
-i \Delta_0^* \eta^y \sigma^y \tau^y & 0
\end{array}
\right) 
\\ & = 
-\eta^y \sigma^y \tau^y |\Delta_0| \omega^x e^{i \phi_0\omega_z}
\\ & = 
-\eta^y \sigma^y \tau^y |\Delta_0| \left[ \omega^x \cos(\phi_0) + \omega^y \sin(\phi_0) \right]
\; ,
\end{split}
\ee
where $-\phi_0$ is the phase of $-i \Delta_0$.
The singlet $p_x+i p_y$ pairing term is
\be\label{H_2}
\begin{split} 
{\mathcal H}_2 & = 
\left(
\begin{array}{cc}
0 & 
-i \sigma^y \tau^y \frac{1}{2} \{ \Delta_1 , ({\vec \eta} \cdot {\nabla}) \eta^y \}
\\
-i \sigma^y \tau^y \frac{1}{2} \{ \Delta_1^* , \eta^y ({\vec \eta} \cdot {\nabla}) \} & 
0
\end{array}
\right) 
\\ & = -\frac{1}{2}
\sigma^y \tau^y \omega^x \eta^y
\{
|\Delta_1| exp^{+i \phi_1 \omega_z} ,
{\hat D}
\}
\; ,
\end{split}
\ee
where $-\phi_1$ is the phase of $-i \Delta_1$.
Finally, the spinless $p_x+i p_y$ pairing term is
\be\label{H_3}
\begin{split} 
{\mathcal H}_3 & = 
\left(
\begin{array}{cc}
0 & 
i \tau^x \frac{1}{2} \{ \Delta_2 , \eta^z ({\vec \eta} \cdot {\nabla}) \eta^y \}
\\
i \tau^x \frac{1}{2} \{ \Delta_2^* , \eta^y ({\vec \eta} \cdot {\nabla}) \eta^z \} & 
0
\end{array}
\right) 
\\ & = \frac{i}{2}
\tau^x \eta^x \omega^x
\{
|\Delta_2| e^{+i \phi_2 \omega^z} ,
{\hat D}
\}
\; ,
\end{split}
\ee
where $-\phi_2$ is the phase of $i \Delta_2$.
Finally, the quadratic correction to the kinetic term is
\be\label{H_4}
\begin{split} 
{\mathcal H}_4 & = 
\frac{v}{4} \tau^z \left( \omega^z \eta^x \left( \partial_x^2 - \partial_y^2 \right) - 2 \eta^y \partial_x \partial_y \right)
\\ & = 
\frac{v}{4} \tau^z {\hat D}^* {\hat D} \omega^z \eta^x
\; .
\end{split}
\ee

This concludes our derivation of the BdG continuum equations, which we will now investigate in a 
variety of geometries, to determine whether topological zero modes appear.

\subsection{Quantization rule}

In Sec.~\ref{vortex_edge_equiv} we pointed out how the momentum quantization evolves when mapping 
between the vortex and edge geometries. We assumed that the Bohr sommerfeld quantization rule
takes on the form of \eqref{Bohr_Sommerfeld2}. In this short subsection we will briefly deduce
what the quantization rule is for the specific cases we consider in this section.

In the effective continuum theory, the Dirac valley spinor degree of freedom is independent
of the real space position degree of freedom. 
Therefore, the only significant difference between the BdG Hamiltonians we deal with here and those dealt with
in Ref.~\onlinecite{Ashvin:2001}, are the appearance of additional spinor structures - the spin, sublattice 
and Dirac valley spinors. Following the semi-classical derivation of Ref.~\onlinecite{Ashvin:2001},
we can use a coherent state representation of spin $\frac{1}{2}$ not only for the Nambu spinor, but also for the 2 other spinors as well. We introduce classical unit vectors for each one of the spinors ${\hat n}$ for the
Nambu spinor, ${\hat h}$ for the sublattice spinor, and ${\hat t}$ for the Dirac valley 
spinor. For the cases where our fermions have spin, there is also a spin degree of freedom, for which we use
${\hat s}$ for the coherent state representation of the spin.
The path integral formulation will include a Berry phase for each one of the unit vectors, so
\be
\begin{split}
S_B & =
\oint {\bf p} \cdot d{\bf x}
+ \frac{\hbar}{2} \oint {\mathcal A}_{\omega}(\hat n) \cdot d{\hat n}
+ \frac{\hbar}{2} \oint {\mathcal A}_{\sigma}(\hat s) \cdot d{\hat s}
\\ &
+ \frac{\hbar}{2} \oint {\mathcal A}_{\eta}(\hat h) \cdot d{\hat h}
+ \frac{\hbar}{2} \oint {\mathcal A}_{\tau}(\hat t) \cdot d{\hat t}
\; .
\end{split}
\ee
for the spinless case, the spin Berry phase term will be absent.

Finally, following Appendix C of Ref.~\onlinecite{Ashvin:2001}, the Bohr-sommerfeld quantization rule will
be of the form $S_B = 2 \pi \left( \ell + \gamma \right) \hbar$ (the $\gamma$ part coming from the order 
parameter phase winding).
We find therefore, that indeed in the cases we consider here,
the Bohr sommerfeld quantization rule takes on the form of \eqref{Bohr_Sommerfeld2}.


\section{SNS junctions}
\label{SNS}

Perhaps the simplest indication of whether topological zero modes can exist in condensate systems 
is when considering SNS (superconducting-normal-superconducting) junctions\cite{Ashvin:2001}.
In this section we will explore the spectrum of states bound to an SNS junction in the various
condensate phases we mentioned in the previous section. This will serve a first step toward 
determining in which of these phases zero modes may appear.

\begin{figure}
	\centering
		\includegraphics[width=2.0in]{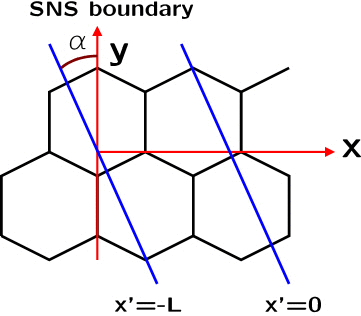}
	\caption{SNS junction on the background of the honeycomb lattice. 
	Our conventions are that the SNS boundaries are at an angle $\alpha$ 
	with the armchair y-direction in our conventions.}
	\label{fig:SNS}
\end{figure}

The SNS junction is modeled as an infinite strip of width $L$ in the continuum limit,
where the pairing function vanishes (see Fig.~\ref{fig:SNS}). On the two sides of the strip are condensate regions,
with a uniform pairing function, with a relative $U(1)$ phase $\phi$.
For the SNS junction we have the pairing function
\be
\Delta(x) = 
\Bigg\{
\begin{array}{ll}
\Delta e^{i \phi} & x < -L \\
0 & -L <x <0 \\
\Delta & 0 < x
\end{array}
\; ,
\ee
where $x$ is the coordinate in the direction perpendicular to the SNS junction walls. 
 
\subsection{Singlet s-wave condensate}
\label{s_wave_SNS}

Combining the kinetic energy \eqref{H_0} and s-wave pairing term \eqref{H_1}
the BdG equation for the s-wave condensate takes the form
\be\label{H_BdG}
\begin{split}
{\mathcal H}_{BdG} \psi = & E \psi \\ = &
\Big[ 
\mu \omega^z - i v {\hat D}
\\ &
- \Delta \left( \omega^x \cos{(\phi)} + \omega^y \sin{(\phi)} \right) \eta^y \tau^y \sigma^y
\Big] \psi
\; ,
\end{split}
\ee
where $\phi$ is the $U(1)$ phase of the order parameter,
and to avoid clutter we have dropped the $0$ subscript from both the phase and order parameter magnitude.
It is convenient to work in the London gauge - we use a unitary transformation
${\mathcal O} = e^{-i \frac{\phi}{2} \omega^z}$ in order to rotate the phase $\phi \rightarrow 0$.

We are free to choose $\psi$ to be eigenstates of $\tau^y, \sigma^y$,
such that $\tau^y \sigma^y \rightarrow \tau \sigma$. This 4-fold degeneracy applies to the
entire quasiparticle energy spectrum. We note here that the explicit appearance of 
$\tau^y, \sigma^y$ is misleading, because the BCS Hamiltonian in this phase is in fact
$SU(2)$ invariant for \emph{both} the spin and Dirac spinors. It is perhaps more appropriate to
write $\sigma^y \tau^y = - \epsilon_{\tau} \epsilon_{\sigma}$ a product of the two totally antisymmetric 
$2 \times 2 $ tensors in the spin and Dirac spinor space. The $SU(2)$ invariance in both these spinor 
spaces then becomes evident (these appear in the pairing terms of the Hamiltonian).
Further assuming that the SNS junction is aligned with a armchair line of the honeycomb lattice (y-direction in our
conventions - see Fig.~\ref{fig:SNS}) and that 
$\psi$ only varies in the x-direction, which we are allowed to assume in a y-infinite system,
we find the BdG Hamiltonian reduces to
\be\label{s_wave_x}
{\mathcal H}_{BdG} =
\Big[ 
\mu \omega^z - i v \eta^x \partial_x 
- \Delta \omega^x \eta^y \tau \sigma
\Big]
\; .
\ee
Apart from the expected (see section~\ref{GenZeroModes}) symmetry
$ \omega^x {\mathcal H}_{BdG} \omega^x = -{\mathcal H}_{BdG}^*$ it is also
easy to show that 
$\eta^x \omega^z {\mathcal H}_{BdG} \eta^x \omega^z = {\mathcal H}_{BdG}$.
This relation is special for y-independent states (once y-variation is allowed this is no longer
a symmetry of the Hamiltonian). The additional symmetry will yield a further double degeneracy
of the spectrum, so we expect every energy level to be 8-fold degenerate. 
The eigenstates of \eqref{s_wave_x} obey
\be
\partial_x \psi = A \psi =
\frac{1}{i v} \eta^x
\left[ 
\mu \omega^z 
- E 
- \Delta \omega^x \eta^y \tau \sigma
\right] \psi
\; .
\ee
The matrix $A$ is then diagonalized, using some (x-independent) similarity transformation $U$,
and we find it has 4 eigenvalues
\be
\begin{split}
U^{-1} \cdot A \cdot U = &
\\
 \frac{1}{v}
\textrm{Diagonal} \Big( &
- \sqrt{\Delta^2 - E^2} + i \mu ,
- \sqrt{\Delta^2 - E^2} - i \mu , \\ &
+ \sqrt{\Delta^2 - E^2} + i \mu , 
+ \sqrt{\Delta^2 - E^2} - i \mu
\Big)
\; .
\end{split}
\ee
The diagonalizing transformation is
\be
U = 
\left(
\begin{array}{llll}
 \alpha +i E  & i E -\alpha  & i E -\alpha  & \alpha +i E  \\
 -\alpha -i E  & i E -\alpha  & \alpha -i E  & \alpha +i E  \\
 \Delta  \sigma  \tau  & -\Delta  \sigma  \tau  & \Delta  \sigma  \tau  & -\Delta  \sigma  \tau  \\
 \Delta  \sigma  \tau  & \Delta  \sigma  \tau  & \Delta  \sigma  \tau  & \Delta  \sigma  \tau 
\end{array}
\right)\; ,
\ee
where $\alpha = \sqrt{\Delta^2 - E^2}$.
When considering energy levels well below the condensate gap $E \ll \Delta$
we have $\sqrt{\Delta^2 - E^2} > 0$.
Therefore, for $x>0$ we have the normalizable solution 
\be
\psi(x) = U \cdot
\left(
e^{\left(- \sqrt{\Delta^2 - E^2} + i \mu \right) \frac{x}{v}} a_1,
e^{\left(- \sqrt{\Delta^2 - E^2} - i \mu \right) \frac{x}{v}} a_2, 0, 0
\right)^T
\; .
\ee
For $x<-L$ we have the normalizable solution
\be
\psi(x) = {\mathcal O} \cdot U \cdot
\left(
0, 0, 
e^{\left(+ \sqrt{\Delta^2 - E^2} + i \mu \right) \frac{x}{v}} b_1,
e^{\left(+ \sqrt{\Delta^2 - E^2} - i \mu \right) \frac{x}{v}} b_2
\right)^T
\; .
\ee
We must now solve for the wavefunction $\psi$ in the normal region,
and then match the wavefunction at the interfaces.
For $-L < x < 0$ the pairing function vanishes, in which case
$A = \frac{1}{i v} \eta^x \left( \mu \omega^z - E \right)$, and 
since all solutions are normalizable in this region, we
can write
\be
\psi(x) = e^{A x} \psi(x=0)
\; .
\ee
Now we match the wavefunction at $x=0,-L$. This will yield a set of linear equations with the variables
$a_{1,2},b_{1,2}$. It can then be recast as a matrix equation $B \cdot (a_1,a_2,b_1,b_2)^T = 0$,
and for a non-trivial solution, we require that $Det(B) = 0$. The equation for the determinant turns out to be
\be\label{s-wave_quantization}
\begin{split} &
\frac{4 \cos \left(\frac{2 L E }{v}\right) E ^2}{\Delta ^2}
+
\frac{4 \sqrt{\Delta ^2-E ^2} \sin\left(\frac{2 L E }{v}\right) E }{\Delta ^2}
\\ &
-
2 \cos \left(\frac{2 L E}{v}\right)
-
2 \cos (\phi) = 0
\; .
\end{split}
\ee

The condition Eq.~\eqref{s-wave_quantization} induces a quantization of the energy values. In 
particular, we can now investigate 
whether zero modes are possible. With $E=0$, Eq.~\eqref{s-wave_quantization} becomes
$ \cos (\phi) = -1 $, which means only when the two superconducting slabs have a $\pi$ phase difference 
does a zero mode eigenstate appear. Furthermore, we can find the spectrum of low energy states - we 
use the limit $E \ll \Delta$ to approximate Eq.~\eqref{s-wave_quantization} as
$\cos \left(\frac{2 L E}{v}\right)
+ \cos (\phi) \approx 0$
which then yields
\be
E \approx \frac{v}{2L} 2 \pi \left( n + \frac{1}{2} - \frac{\phi}{2 \pi} \right)
\; .
\ee
We find the spectrum is evenly spaced, with a spacing $\delta E = \frac{v}{2L} 2 \pi$.

The zero mode wavefunctions can also be found (from the null space of the matrix $B$, 
with $E=0$ and $\phi = \pi$).
Including the $\tau$ and $\sigma$ spinors we have been ignoring, 
$\chi_{\tau} = \left(1 , i \tau \right)^T$ and
$\chi_{\sigma} = \left(1 , i \sigma \right)^T$
we find a total of 8 solutions
\be
\psi_{\eta \sigma \tau}(x) = \psi_{\eta}(0) \otimes 
\chi_{\tau}
\otimes
\chi_{\sigma}
e^{ \eta \frac{i x \mu }{v}} 
\Bigg\{
\begin{array}{cc}
e^{+\frac{(x+L) \Delta }{v}} & x < -L \\
1 & -L < x < 0 \\
e^{-\frac{x \Delta }{v}} & x>0
\end{array}
\; ,
\ee
where $\psi_{\eta}(0) = {\mathcal N} \left( \sigma \tau , - \eta \sigma \tau ,1,\eta \right)^T$,
$\eta =\pm 1$ and ${\mathcal N}$ is a normalization factor. Here, and elsewhere in this section, 
the column vector $\psi_{\eta}(0)$ has the entries $\left( u_1, u_2, v_1, v_2 \right)^T$, 
where $1,2$ are the sublattice indices. These two solutions are independent 
(even at half filling $\mu = 0$), and in fact are also eigenstates of 
$\eta^z \omega^x \psi_{\eta \sigma \tau} = \sigma \tau \psi_{\eta \sigma \tau}$ and 
$\eta^x \omega^z \psi_{\eta \sigma \tau} = -\eta \psi_{\eta \sigma \tau}$ .
It is also evident that $\omega^x \psi_{\eta \sigma \tau}^* = \sigma \tau \psi_{-\eta, -\sigma, -\tau}$.

As mentioned in section~\ref{ContinuumLimit}, the honeycomb lattice tight binding model in the continuum limit
includes a quadratic derivative correction to the Dirac operator \eqref{H_4}.
Since the degeneracy in $\tau = \pm 1$ stems from the $SU(2)$ valley spinor symmetry,
which is explicitly broken (and reduced to $U(1)$) by \eqref{H_4},
we should investigate whether this term splits the 8 zero modes we have found.
This will be our first step in exploring whether the zero modes appear in the 
original lattice model.
Since we are considering here y-independent states, the correction reduces to 
\be\label{x_correction}
{\mathcal H}_4 =
\frac{v}{4} \tau^z \omega^z \eta^x \partial_x^2
\; .
\ee
The correction commutes with $\eta^x \omega^z$, and is spin $SU(2)$ invariant,
so the quantum numbers $\eta$ and $\sigma$ are conserved, so only $\tau$ can mix
and a quadruple degeneracy of every energy level will still hold. 
A simple calculation yields that all the matrix elements between the zero modes
induced by the correction vanish, and so this term does not split the zero modes,
in this SNS geometry, with the armchair alignment.


So far we have only considered a very particular alignment of the SNS junction walls - the y-direction in our 
conventions for the honeycomb lattice. 
Now we turn to investigate whether different orientations
of the SNS junction behave different. Rotating the SNS junction counterclockwise by an angle $\alpha$
(see Fig.~\ref{fig:SNS}),
and assuming the eigenstates only vary in the direction perpendicular to the SNS junction walls,
which we denote by $x'$, the only term that changes in \eqref{H_BdG} in the London gauge is
\be\label{SNS_Dirac_op}
{\hat D}(\alpha) = \left[ \cos{(\alpha)} \eta^x  + \sin{(\alpha)} \omega^z \eta^y \right] \partial_{x'} = 
\eta^x e^{+ i \alpha \omega^z \eta^z} \partial_{x'}
\; .
\ee
It is easy to show that the unitary transformation 
$U(\alpha) = e^{- i \frac{\alpha}{2} \omega^z \eta^z }$ will rotate $\alpha \rightarrow 0$,
mapping this problem directly onto the problem with the SNS junction parallel to the y-direction
$U(\alpha)^{\dagger} {\mathcal H}_{BdG}(\alpha) U(\alpha)^{\phantom\dagger} = {\mathcal H}_{BdG}(\alpha=0)$.
Furthermore, the symmetry 
$\left[{\mathcal H} , \omega^z \eta^x \right] = 0$
will simply be replaced by $\left[{\mathcal H}(\alpha) , U(\alpha)^{\phantom\dagger} \omega^z \eta^x U(\alpha)^{\dagger} \right] = 0$.
Thus, we conclude that the eigenvalue spectrum, at least when ignoring the quadratic correction to 
the kinetic energy, is completely independent of the SNS junction orientation, and will remain 8-fold degenerate.

The quadratic correction to the kinetic energy with the SNS junction walls rotated, takes the form
\be
\begin{split}
{\mathcal H}_4 = &
\frac{v}{4} \tau^z \left( {\hat D(\alpha)}^* {\hat D(\alpha)} \right) \omega^z \eta^x
\\ = &
\frac{v}{4} \tau^z e^{i 2 \alpha \omega^z \eta^z} \omega^z \eta^x \partial_{x'}^2
\; .
\end{split}
\ee
Now we want to examine how the correction transforms under the unitary transformation that 
rotates $\alpha \rightarrow 0$ in ${\mathcal H}_{BdG}(\alpha)$. We note first that all terms in 
${\mathcal H}_4$ apart from the operator $\eta^x$ remain invariant under this unitary transformation.
The correction becomes
\be\label{splitter2}
\begin{split}
U^{\dagger} {\mathcal H}_4 U = &
\frac{v}{4} \tau^z e^{i 2 \alpha \omega^z \eta^z} \omega^z U^{\dagger} \eta^x U \partial_{x'}^2 
\\ = &
\frac{v}{4} \tau^z \left[
\cos{(3 \alpha)} \eta^x \omega^z -
\sin{(3 \alpha)}\eta^y 
\right] \partial_{x'}^2
\; .
\end{split}
\ee
The emergence of the $3 \alpha$ factors may seem a bit surprising, but in fact this is
a consequence of the underlying 3-fold rotation symmetry of the honeycomb lattice - the splitting
is the same if we rotate the SNS junction by $2\pi/3$.
The correction naturally reduces to Eq.~\eqref{x_correction} when $\alpha =0$.
In fact, the first term above is precisely the $\alpha=0$ splitting matrix
multiplied by the factor $\cos{(3 \alpha)}$.
Using this fact, we need only compute the matrix elements of the second term when 
projected onto the subspace of zero modes. The result is 
\be\label{SNS_splitting}
\begin{split} &
\bra{\psi_{\eta' \sigma' \tau'}} U^{\dagger} {\mathcal H}_4 U \ket{\psi_{\eta \sigma \tau}} 
= - \sin{(3 \alpha)} \delta_{\sigma' \sigma} \eta^y_{\eta' \eta} \frac{\Delta ^2}{8 (v+L \Delta )}
\\ &
\left[
\tau^x_{\tau' \tau} \sin{\left( \frac{2 L \mu}{v} \right)}
+
\tau^y_{\tau' \tau} \left( 1 + \cos{\left( \frac{2 L \mu}{v} \right)} \right)
\right]
\; ,
\end{split}
\ee
In general the eigenvalues of this matrix will be non-vanishing (except for the special pathological cases 
$\frac{2 L \mu }{v} = \pi$ and $\alpha$ an integer multiple of $\pi/3$), thus splitting the zero mode energies. 

To conclude this subsection, we have shown that in the SNS junction geometry, 
the honeycomb Dirac dispersion s-wave condensate can 
support topological zero modes, when an odd phase winding is present ( the $\pi$ phase difference between the condensate slabs). However, these zero modes split when we take into account quadratic corrections to the kinetic energy, which are intrinsically present in the honeycomb lattice. Only in the case where the junction walls are 
aligned as armchair boundaries in the honeycomb lattice ($\alpha = 0$), do the zero modes remain unsplit
by the quadratic correction.

\subsection{Singlet p+ip condensate}

Now we turn to SNS junctions in the $p_x + i p_y$ singlet phase.
Our analysis will be very similar to that carried out in the previous subsection 
for the s-wave phase, and as such we will describe our calculations in much less detail. 
Here and throughout the remainder of our manuscript, we will assume $\mu >0$ for all
of the $p_x + i p_y$ phases, since these are gapped only away from half filling, and since
essentially the same result can be found for $\mu <0$ (due to the honeycomb particle-hole 
symmetry).

With a (piecewise) uniform pairing function, combining the kinetic energy 
\eqref{H_0} and the spin-singlet $p_x+i p_y$ pairing term \eqref{H_2}
the BdG equation takes the form
\be\label{p_wave_BdG}
\begin{split} &
{\mathcal H}_{BdG} \psi = E \psi \\ = &
\Big[ 
\mu \omega^z - i v {\hat D}
- \Delta \left( \omega^x \cos{(\phi)} + \omega^y \sin{(\phi)} \right) \eta^y
{\hat D}
 \tau^y \sigma^y
\Big] \psi
\; ,
\end{split}
\ee
where we have dropped the $1$ subscript from both the order parameter phase and magnitude, to avoid clutter.
As before we will work in the London gauge - the unitary transformation
${\mathcal O} = e^{-i \frac{\phi}{2} \omega^z}$ will rotate the phase $\phi \rightarrow 0$.

Given what we have learned about the significance of the SNS junction orientation in the previous subsection,
we begin by briefly addressing this point. As in the s-wave case, we assume the angle between the SNS 
boundaries and the y-axis is $\alpha$, and consider eigenstates with spatial variation only in the 
direction perpendicular to the SNS boundaries. 
We use the same unitary transformation $U(\alpha)$ to rotate $\alpha \rightarrow 0$ in the kinetic energy.
The pairing function now includes ${\hat D}$, which is also rotated to it's $\alpha = 0$ value, and all 
the other operators remain invariant.
Choosing in addition eigenstates of $\tau^y, \sigma^y$,
the BdG Hamiltonian then reduces to
\be\label{p_wave_x}
{\mathcal H}_{BdG} =
\Big[ 
\mu \omega^z - i v \eta^x \partial_x 
+ i \Delta \omega^x \eta^z \tau \sigma \partial_x
\Big]
\; .
\ee
At this point we will note, that the BdG Hamiltonian we obtain here, just as its s-wave counterpart, 
has an extra symmetry $\left[ {\mathcal H}_{BdG}, \eta^x \omega^z \right] = 0$.
As a result, we expect the spectrum to be 8-fold degenerate.

Following the same procedure elaborated in the previous subsection, we find the energy quantization 
condition in the SNS junction to be
\be
\begin{split}
0 = & 
\frac{2}{\Delta^2 (E - \mu)^2} 
\Big[
-\left(E ^2-\mu ^2\right) \cos (\phi ) \Delta ^2
\\ &
-
\left(\left(E ^2-\mu ^2\right) \Delta ^2+2v^2 E ^2\right) \cos \left(\frac{2 L E }{v}\right)
\\ &
+2 i v E  \sqrt{\left(E ^2-\mu^2\right) \Delta ^2+v^2 E ^2} 
\sin \left(\frac{2 L E }{v}\right)
\Big]
\; .
\end{split}
\ee
For low energies $E \ll \Delta$ this reduces to $\cos{(\phi)} + \cos{\left(\frac{2 L E }{v}\right)}$ 
and we obtain the low energy spectrum
\be
E = \frac{v \pi }{L} \left( n + \frac{1}{2} - \frac{\phi}{2 \pi}\right)
\; ,
\ee
identical to the spectrum we found for the s-wave phase, and including zero modes, only when $\phi = \pi$.

Using the same conventions we used for the s-wave case,
the zero mode wavefunctions we find, when taking $\phi = \pi$, are
\be
\psi_{\eta \sigma \tau}(x) = \psi_{\eta}(0) \otimes 
\chi_{\tau}
\otimes
\chi_{\sigma}
e^{ \eta \frac{i x \mu }{v}} 
\Bigg\{
\begin{array}{cc}
e^{\frac{(L+x) \Delta  \mu }{v (v + i \eta \Delta ) }} & x < -L \\
1 & -L < x < 0 \\
e^{-\frac{x \Delta  \mu }{v (v - i \eta \Delta ) }} & x>0
\end{array}
\; ,
\ee
where $\psi_{\eta}(0) = {\mathcal N} \left( 
-i \sigma  \tau , i \eta  \sigma  \tau ,1, \eta
\right)^T$
and ${\mathcal N}$ is a normalization factor.
Quite similarly to the s-wave zero modes, these solutions are eigenstates of
$\eta^x \omega^z \psi_{\eta \sigma \tau} = - \eta \psi_{\eta \sigma \tau}$ and of
$\eta^z \omega^y \psi_{\eta \sigma \tau} = \sigma \tau \psi_{\eta \sigma \tau}$.
From the form of the solutions it is also clear that 
$\omega \left( \psi_{\eta \sigma \tau} \right)^* = i \sigma \tau \psi_{-\eta ,-\sigma ,-\tau} $.

Finally we discuss the influence of the quadratic correction to the kinetic energy.
Here the value of $\alpha$ is significant, and so we go straight to \eqref{splitter2},
and calculate the matrix elements in the zero mode subspace. We find that all the matrix elements 
\emph{vanish}, and so zero modes are not split to first order in the correction.

\subsection{SNS junctions summary}

To conclude this section exploring the SNS junction geometry,
we recap the results of our calculations.
For simplicity we have limited our discussion to wavefunctions 
uniform in the direction parallel to the walls.
The spin-singlet s-wave phase supports zero modes only when $\alpha = 0$
(armchair boundary), and otherwise does not possess zero modes, with splitting 
due to the quadratic correction to the kinetic energy. 
The spin-singlet $p_x + i p_y$ phase supports zero
modes (to first order in perturbation theory in the quadratic correction).

\section{Edge states}
\label{Edge_states}

In this section we will investigate the edge state spectrum
of the various phases we are exploring in this article.
For convenience we will consider an edge where the honeycomb 
lattice abruptly ends, and assume the pairing function 
is uniform in space. 
We expect the bound states with low energy to appear
with low momentum in the direction parallel to the edge (in the lattice model),
and because of this one needs to be somewhat cautious when thinking about the continuum limit.
For the armchair edge of the honeycomb lattice,
the Dirac point momenta are perpendicular to the boundary, and so 
low transverse momentum can be well described even in the continuum limit.
For a zigzag edge, the Dirac momenta ${\bf Q}$ are parallel to the edge, and so small 
momentum in the lattice model ${\bf p} = {\bf Q} + {\bf q} \approx 0 $ parallel to this edge,
corresponds to momentum of order the Dirac momentum in the continuum theory 
${\bf q} \approx -{\bf Q}$. Under these extreme conditions, the validity of the continuum 
limit approximation for the lattice model breaks down - the real momentum is quite far away from the 
Dirac point. We will therefore explore only the armchair edge in our present work (corresponding to 
$\alpha = 0$ in the previous section and shown in Fig.~\ref{fig:SNS}).

\subsection{Boundary conditions in the continuum limit of the honeycomb lattice}\label{boundary_conds}

Since we are taking a continuum limit of lattice models on the honeycomb,
we must study with some care how the boundary conditions must be taken in the 
continuum limit. 

With our choice of the armchair edge, the boundary condition 
of the lattice wavefunction is that it must vanish on some line. The eigenstates of the 
system are in general the Bogoliubov quasiparticles, with creation operators
\be
\gamma^{\dagger} = \sum_{\mu \alpha {\bf r}} \left[
{\tilde u}_{\mu \alpha}({\bf r}) f_{\mu \alpha}^{\dagger}({\bf r}) + 
{\tilde v}_{\mu \alpha}({\bf r}) f_{\mu \alpha}^{\phantom\dagger}({\bf r})
\right]
\; ,
\ee
and the boundary condition corresponds to ${\tilde u} = {\tilde v} = 0$ at the system edge.
We note here that our description applies to non-condensate system as well, by simply taking ${\tilde v}=0$
\emph{everywhere}, in which case the Bogoliubov quasiparticles simply become modes of the fermi gas.

In the continuum limit we employ here, the Bogoliubov quasiparticles are
\be
\gamma^{\dagger} = \sum_{\mu \alpha A} \int_{\bf r} \left[
u_{\mu \alpha A}({\bf r}) \psi_{\mu \alpha A}^{\dagger}({\bf r}) + 
v_{\mu \alpha A}({\bf r}) \psi_{\mu \alpha A}^{\phantom\dagger}({\bf r})
\right]
\; .
\ee
Changing to momentum space in both cases, and using the transformation Eq.~\eqref{continuum}, we identify
\be
\begin{split} &
u_{\mu \alpha R}({\bf q}) = {\tilde u}_{\mu \alpha}({\bf q + Q}) \Pi
\\ &
u_{\mu \alpha L}({\bf q}) = -i \eta^y_{\mu \nu} {\tilde u}_{\nu \alpha}({\bf q - Q}) \Pi
\\ &
v_{\mu \alpha R}({\bf q}) = {\tilde u}_{\mu \alpha}({\bf q + Q}) \Pi
\\ &
v_{\mu \alpha L}({\bf q}) = +i \eta^y_{\mu \nu} {\tilde u}_{\nu \alpha}({\bf q - Q}) \Pi
\; .
\end{split}
\ee
Using these relations we can identify the continuum limit approximation of ${\tilde u}, {\tilde v}$.
We find
\be
\begin{split} &
{\tilde u}_{\mu \alpha}({\bf r}) \approx \Pi \left[
u_{\mu \alpha R} e^{+ i {\bf Q} \cdot {\bf r}}
+ i \eta^y_{\mu \nu}
u_{\nu \alpha L} e^{- i {\bf Q} \cdot {\bf r}} 
\right]
\\ &
{\tilde v}_{\mu \alpha}({\bf r}) \approx \Pi \left[
v_{\mu \alpha R} e^{+ i {\bf Q} \cdot {\bf r}}
- i \eta^y_{\mu \nu}
v_{\nu \alpha L} e^{- i {\bf Q} \cdot {\bf r}}
\right]
\; .
\end{split}
\ee
Next we will use this continuum approximation of the lattice wavefunction
to explore how the boundary conditions translate in the continuum limit.

In our convention, the armchair edge can occur at the line $x=0$, and since we have
${\bf Q} = Q {\hat x}$, the lattice wavefunction boundary condition translates into 
the condition
\be
\begin{split} &
u_R = - i \eta^y u_L 
\\ &
v_R = + i \eta^y v_L 
\; ,
\end{split}
\ee
for the continuum wavefunction, at $x=0$.
Using the various Pauli matrix sets we have defined earlier in this manuscript,
we can reorganize these conditions into the simple form
\be\label{bound_conds}
\omega^z \eta^y \tau^y \psi = \psi
\; ,
\ee
where $\psi=(u,v)$ ( the continuum limit wavefunction).
In what follows, we will assume the lattice occupies the $x>0$ semi-infinite plain,
and use the boundary condition we have derived here.
It is worth while noting that when dealing with non-condensate wavefunctions, where
$\omega_z \psi = \pm \psi$, the boundary condition we have derived here simply reduces to the
previously derived armchair boundary condition\cite{Brey:prb2006}  
in the conventions of Ref.~\onlinecite{akhmerov:085423} $\eta^y \tau^y \psi = \pm \psi$.

\subsection{Singlet s-wave condensate}

We now turn to explore the edge states in the honeycomb 
spin-singlet condensate s-wave phase. Starting from the 
BdG equations Eq.~\eqref{H_BdG}, we choose the order 
parameter phase $\phi=0$, and solutions that are eigenstates of 
$\sigma^y \psi = \sigma \psi$ and $\tau^y \psi = \tau \psi$, in which case
\be
\left[ 
\mu \omega^z - i v \left( \eta^x \partial_x + \omega^z \eta^y \partial_y \right)
-\omega^x \eta^y \tau \sigma \Delta
\right] \psi = E \psi
\; .
\ee
In the semi-infinite geometry, the system is still translationally invariant in the y-direction, 
so we choose solutions of the form $\psi(x,y) = e^{i q y} \psi(x)$.
The BdG equations then become
\be
\left[ 
\mu \omega^z - i v \eta^x \partial_x  + v q \omega^z \eta^y
- \omega^x \eta^y \tau \sigma \Delta
\right] \psi(x) = E \psi(x)
\; .
\ee
The boundary condition \eqref{bound_conds} then requires
$\omega^z \eta^y \psi(x=0) = \tau \psi(x=0)$. 

The set of coupled ODEs can be solved in a manner very similar to the way we solved for the 
$x>0$ region of the SNS junction. we cast the equations in the form $\partial_x \psi = A \psi$,
with the matrix $A$ being independent of $x$. We diagonalize the matrix $A$ with a similarity 
transformation that is x-independent, and then keep those solutions that are exponentially decaying
in $x>0$. In contrast to the SNS junction case, here we allow for a transverse momentum, and for 
this reason the calculations are somewhat more involved. These solutions can be written as 
\be
\psi_{\eta, \tau, \sigma} = 
\chi_{\sigma} \otimes \chi_{\tau} \otimes
e^{-\frac{x F_{\eta}}{v} + i q y}
\left(
\begin{array}{l}
 \frac{\eta  \left(\Delta ^2+i \eta  (B+i E  \eta ) (E -\mu )\right) \sigma  \tau }{2 \Delta  (B-i \eta  \mu )}
   \\
 -\frac{(B+i E  \eta ) \sigma  \tau  \left(q v+F_{\eta }\right)}{2 \Delta  (B-i \eta  \mu )} \\
 \frac{\eta  \left(q v+F_{\eta }\right)}{2 (B-i \eta  \mu )} \\
 \frac{1}{2}
\end{array}
\right)
\; ,
\ee
where we have introduced $B = \sqrt{\Delta^2 - E^2}$
and $F_{\eta} = \sqrt{(B - i \eta \mu)^2 + (q v)^2}$,
and $\eta = \pm 1$. The components of the 4-vector above correspond to the
wavefunction amplitudes $\left( u_1, u_2, v_1, v_2 \right)^T$,
where $1,2$ are the two sublattice indices. Throughout this section
all 4-component vectors will follow this convention.

Note that for small energy $E \ll \Delta$ we have $B \approx \Delta$,
and then further assuming that the momentum is small $q v \ll \Delta$
yields $F_{\eta} \approx (\Delta - i \eta \mu)$, which gives the decay length scale 
we found for the SNS junctions (as well as for the vortex core case, as we will 
see in the next section). At this level, before we impose the edge boundary conditions, 
we find we have 8 solutions per energy $E$ and transverse momentum $q$.

The general spin-eigenvalue solution that decays exponentially in $x>0$ is
$\psi = \psi_{+1} a_{+1} + \psi_{-1} a_{-1}$.
Now we will impose the armchair wall boundary conditions \eqref{bound_conds} to this solution.
The boundary conditions gives 2 linearly independent equations in the variables
$a_{\eta}$. These equations can be cast in matrix form, and for a non-trivial 
solution $a_{\eta} \neq 0$ to exist, the determinant of the matrix must vanish.
The resultant equation for the determinant, is a quantization condition for the energies $E$.
The precise form of this quantization rule is 
\be
\begin{split}
0 = & 
i \tau  B^3 + E  \mu  \left(F_{-1}-F_1\right)
\\ - &
i B \Big(
q^2 \tau v^2+q \left(\tau \left(F_{-1}+F_1\right)-2 E \right) v
\\ - &
E \left(F_{-1}+F_1\right)+\tau  \left(F_{-1} F_1-\mu ^2\right)
\Big)
\; .
\end{split}
\ee

Considering low energies $E \ll \Delta$, and small momentum $q v \ll \Delta$, we can approximate
the quantization condition to 
$ \left((E -q v \tau ) \Delta ^2+E  \mu ^2\right) \approx 0 $ 
which yields
\be\label{s_wave_edge_spectrum}
E \approx \tau \frac{q v \Delta ^2}{\Delta ^2+\mu ^2}
\; .
\ee
We find that zero modes exist for $q=0$. Next we will obtain the zero mode wavefunctions, by taking 
$q=0,E=0$, we recover the amplitudes
$ \left( a_{+1},a_{-1} \right) = \left( i-\tau ,\tau +i  \right) $.
The complete wavefunctions of the zero modes are
\be
\begin{split}
\psi^0_{\sigma \tau} = &
\chi_{\sigma} \otimes \chi_{\tau} \otimes
e^{-x \frac{\Delta}{v} } 
\\ &
\left[
i (1+i\tau) \psi_0 e^{+ i \frac{\mu}{v} x}
-i (1-i\tau) \eta^z \psi_0^* e^{- i \frac{\mu}{v} x}
\right]
\; ,
\end{split}
\ee 
where
$\psi_0 = \left( \sigma \tau, -\sigma \tau, 1, 1\right)^T$.
We find a total of 4 zero modes ($\sigma, \tau = \pm 1$). 

The particle-hole relation of the BCS Hamiltonians that given
an eigenstate $\psi_E$ with energy $E$, $\omega^x \psi_E^*$ is 
also an eigenstate with energy $-E$, when applied to the 4 zero 
modes we find here, gives 4 states that are \emph{orthogonal}
to the zero modes we found. This surprising result is understood when considering how the 
boundary condition behaves. Starting from $\omega^z \eta^y \tau^y \psi = \psi$, we want to know 
what boundary condition is satisfied by ${\overline \psi} = \omega^x \psi^*$. It can be easily
shown that the boundary condition is $\omega^z \eta^y \tau^y {\overline \psi} = - {\overline \psi}$.
This result shows us that the ${\overline \psi}$ states are \emph{precisely} the ones discarded
by the boundary condition in this case! Therefore, the only states satisfying the 
boundary conditions are the 4 zero modes we have found above. Finally, it is amusing to mention
another consequence of these boundary conditions - the superposition yielding a Majorana fermion
$\psi + {\overline \psi}$, \emph{cannot} be taken here! Therefore, despite the existence of zero 
modes, they cannot form Majorana fermion states.

The edge states energy spectrum at low momentum $q$ is linear in the momentum,
and we now proceed to briefly calculate the approximate edge state wavefunctions for 
these low energies. The boundary conditions, cast as linear equations in the 
coefficients $a_{\eta}$, can be linearized in energy and momentum. In this case the 
energy quantization condition we derive is
\be
E = \frac{\Delta q \tau  v \left(q v + 2 \Delta \right)}{ \Delta \left( q v + 2 \Delta \right) + 2\mu^2 - q v \mu \tau }
\; ,
\ee
with $\omega = \pm 1$. Linearizing in momentum $q$, this result reduces to \eqref{s_wave_edge_spectrum}.
The solution for $a_{\eta}$ we find with this value of $E$ is
$a_{+1} = a_{-1}^* = (\Delta -i \mu ) (q v+(\Delta +i \mu ) (i \tau +1))$.
With these coefficients, linearizing everything in momentum $q$ yields
\be
\psi_{\sigma, \tau} = 
\chi_{\sigma} \otimes \chi_{\tau} \otimes
e^{-x \frac{\Delta}{v} + i q y}
\Bigg[ e^{+ i \frac{\mu}{v} x} \Psi a_{+1} + e^{- i \frac{\mu}{v} x} \eta^z \Psi^* a_{+1}^* \Bigg]
\; ,
\ee
where 
\be
\Psi = 
\left(
\begin{array}{l}
 \left(\Delta ^2+i (\Delta +i E ) (E -\mu )\right) \sigma  \tau  \\
 -(\Delta +i E ) (q v+\Delta -i \mu ) \sigma  \tau  \\
 \Delta  (q v+\Delta -i \mu ) \\
 \Delta  (\Delta -i \mu )
\end{array}
\right).
\ee
We find that every energy level has a 4-fold degeneracy, including the zero modes.
This is the minimal expected degeneracy, required by the $SU(2)$ symmetries of both
the spin and the valley spinor. Since the zero modes posses only this minimal 
degeneracy as well, it is topologically protected.

Considering the quadratic correction to the kinetic energy term, since the zero modes only vary in 
the x-direction, the splitting term reduces to \eqref{x_correction}
A straightforward calculation then shows that all the matrix elements between the zero mode wavefunctions
$\psi^0_{\sigma \tau}$ vanish, and we find that there is no splitting to first order.
This is in agreement with subsection~\ref{s_wave_SNS} because the armchair edge corresponds to the angle $\alpha = 0$ in \eqref{SNS_splitting}, yielding no splitting.

\subsection{Singlet p+ip condensate}

In this subsection, we turn to explore the edge states in the honeycomb 
spin-singlet condensate $p_x + i p_y$ phase. Starting from the 
BdG equations Eq.~\eqref{p_wave_BdG}, 
we proceed with a calculation that is only slightly different than that performed for 
the s-wave phase in the previous subsection.

The system is translationally invariant in the y-direction, 
so we choose solutions of the form $\psi(x,y) = e^{i q y} \psi(x)$. Furthermore,
we choose the order parameter phase $\phi=0$, and solutions that are eigenstates of 
$\sigma^y \psi = \sigma \psi$ and $\tau^y \psi = \tau \psi$. All this yields
\be
\Big[ 
\mu \omega^z - i v {\hat D}
- \Delta \omega^x \eta^y
{\hat D}
 \tau \sigma
\Big] \psi = E \psi
\; ,
\ee
where ${\hat D} =
\left( \eta^x \partial_x + \omega^z \eta^y i q \right) $.
And the boundary condition \eqref{bound_conds} then requires
$\omega^z \eta^y \psi(x=0) = \tau \psi(x=0)$. 

We find the solutions to this set of coupled ODEs in the same manner
as in the previous subsection. We find those solutions that are exponentially 
decaying in $x>0$, and linearize them in momentum and energy, expecting a 
low energy relation $E \sim q$. We find the general (linearized) solution
\be
\begin{split} &
\psi_{\eta \sigma \tau} = 
\chi_{\sigma} \otimes \chi_{\tau} \otimes
e^{\frac{i x \eta  \mu }{v-i \Delta  \eta }}
\\ &
\left(
\begin{array}{l}
 i (i v E +\Delta  \eta  (E +\mu )) \sigma  \tau  \\
 (v E  \eta +q \Delta  (v-i \Delta  \eta ) \eta -i \Delta  (E +\mu )) \sigma  \tau  \\
 \Delta  \eta  (q (\Delta +i v \eta )+\mu ) \\
 \Delta  \mu 
\end{array}
\right)
\; ,
\end{split}
\ee
where $\eta = \pm 1$. We take the general solution $\psi = \sum_{\eta} \psi_{\eta} a_{\eta}$
and find which coefficients $a_{\eta}$ will satisfy the armchair boundary conditions. 
The set of equations for $a_{\eta}$ can be cast in a matrix form, and for a non-trivial solution
to exist, the matrix determinant must vanish. This yields the approximate quantization condition
for the low energy spectrum
\be
\begin{split} &
2 \Delta \sigma \left(
\Delta \left( v^2 + \Delta^2 \right) q^2 
+ \left(E v^2 + \Delta^2 (E + 2 \mu) \right) q 
- 2 v E \mu \tau 
\right) 
\\ &
= 0
\; ,
\end{split}
\ee
with the solutions
\be
E = -\frac{\Delta  \left(q^2 v^2+q^2 \Delta ^2+2 q \Delta  \mu \right)}{q v^2-2 \mu  \tau  v+q \Delta ^2} \approx 
\frac{\Delta ^2 \tau  q}{v}+O\left(q^2\right)
\; .
\ee
As expected, we indeed find a branch of low-energy states with energy linear in the transverse momentum,
and zero modes for $q=0$. Next, we find the approximate edge state wavefunctions. The solutions
for the coefficients are
\be
a_{\eta} = q (\Delta - \eta i v)+\mu + i \mu \eta \tau
\; .
\ee
The full edge state wavefunctions we find then, after linearizing them with respect to the momentum $q$, are
\be
\begin{split} &
\psi_{\sigma \tau} = 
\chi_{\sigma} \otimes \chi_{\tau} \otimes
e^{-\frac{x \Delta  \mu }{v^2+\Delta ^2} + i q y}
\\ &
\Bigg[
\cos \left(\frac{v x \mu }{v^2+\Delta ^2}\right)
\left(
\begin{array}{l}
 \sigma  \left(q \left(\tau  v^2-\Delta  v-\Delta ^2 \tau \right)-v \mu \right) \\
 i \sigma  \left(q \left(v^2-\Delta  \tau  v-\Delta ^2\right)-v \mu  \tau \right) \\
 i v (q \Delta +\mu ) \tau  \\
 v (q \Delta +\mu )
\end{array}
\right)
\\ &
+ 
\sin \left(\frac{v x \mu }{v^2+\Delta ^2}\right)
\left(
\begin{array}{l}
 -\sigma  \left(q \Delta ^2+v \mu  \tau \right) \\
 i \sigma  (v \mu +q \Delta  (2 v+\Delta  \tau )) \\
 -i v (-2 q \Delta -\mu +q v \tau ) \\
 v (q v-\mu  \tau )
\end{array}
\right)
\Bigg]
\; .
\end{split}
\ee
We find every low-energy state is 4-fold degenerate - for $E \neq 0$,
there is spin degeneracy, and a 2-fold degeneracy of the product $\tau q$.
For the zero modes ($q = 0$) there is still a 4-fold degeneracy, in both spin
and valley spinor degeneracy. As for the s-wave case, this is the minimal 
degeneracy of each energy level, and as such, the zero modes are topologically protected.

Finally, we turn to examine what influence the quadratic correction to the kinetic energy 
has over the zero modes, since this perturbation breaks the valley spinor $SU(2)$ symmetry.
As in the s-wave case, the zero modes have no y-dependence, so the correction reduces to 
\eqref{x_correction}. 
A straightforward calculation of the matrix elements between 
the different zero modes, yields these all vanish, and so there is no splitting from this correction,
to first order.

\section{Vortex core zero mode bound states}
\label{Vortex_states}

In this section we will investigate whether zero mode bound states at vortex cores exist,
in various phases.

\subsection{Singlet s-wave condensate}
\label{s_wave}

The simple s-wave singlet-pairing condensate phase on the honeycomb lattice, 
has an eigenvalue spectrum determined by the BdG equation
\be\label{BdG3}
\left( {\mathcal H}_0 + {\mathcal H}_1 \right)\psi = E \psi
\; ,
\ee
where we refer to \eqref{H_0} and \eqref{H_1} for the full details of the kinetic and pairing term.
In this section we will find the zero-mode solutions in \eqref{BdG3} explicitly.
As opposed to the calculation of Ref.~\onlinecite{Ghaemi:2007}, we allow for a non-zero chemical potential
(corresponding to slight deviations from half filling), and we find \emph{exact} solutions for the zero modes. 

As a first step, we choose solutions that are $\sigma^y \tau^y$ eigenstates, precisely as in the SNS 
and edge geometries,
and as a result the energy spectrum will be at least 4-fold degenerate.
We will model the vortex by assuming the form $-i \Delta_0 = \Delta(r) e^{i \phi}$, with $\Delta(r)$ real. 

We begin by considering the half filling case ($\mu = 0$). The BdG equation then become
\be
\left(
\begin{array}{cc}
 -i v \left({\vec \eta} \cdot \nabla \right) & \Delta(r) e^{i \phi} \eta^y \sigma \tau \\
\Delta(r) e^{-i \phi} \eta^y \sigma \tau & - i v \left({\vec \eta}^* \cdot \nabla \right)
\end{array}
\right)
\psi = 0
\; .
\ee
In this special case, one can obtain solutions that exist on only one of the two sublattices.
this becomes obvious when multiplying the equation set ${\mathcal H} \psi = 0$ by $\eta^y$
on the left, resulting in
\be
\left(
\begin{array}{cc}
 -i v \left(-i \eta^z \partial_x + \partial_y \right) & \Delta(r) e^{i \phi} \sigma \tau \\
\Delta(r) e^{-i \phi} \sigma \tau & - i v \left(-i \eta^z \partial_x - \partial_y \right)
\end{array}
\right)
\psi = 0
\; .
\ee
We can then choose to consider a solution on one of the sublattices $1,2$ in which case 
$\eta^z \rightarrow \eta = \pm 1$.
Writing the equations in polar coordinates
\be
\left(
\begin{array}{cc}
- v \eta e^{+i \eta \phi} \left( \partial_r + \frac{i \eta}{r} \partial_{\phi} \right) & 
\Delta(r) e^{+i \phi} \sigma \tau \\
\Delta(r) e^{-i \phi} \sigma \tau & 
- v \eta e^{- i \eta \phi} \left(\partial_r - \frac{i \eta}{r} \partial_{\phi} \right)
\end{array}
\right)
\psi = 0
\; .
\ee
We observe that a for $\eta = +1$, we can try a solution $\psi = (u,v)$ where $u,v$ are 
independent of the angle $\phi$, and for $\eta = -1$, we can try a solution of the 
form $\psi = \frac{1}{r} (e^{i \phi} u , e^{-i \phi} v)$. For both choices, the equations reduce to 
\be
\left(
\begin{array}{cc}
- v \partial_r & 
\Delta(r) \eta \sigma \tau \\
\Delta(r) \eta \sigma \tau & 
- v \partial_r
\end{array}
\right)
\left( 
\begin{array}{c}
u \\ v
\end{array}
\right) = 0
\; .
\ee
Rewriting these equations using the Nambu pauli matrices
\be
\left[ - v \partial_r + \Delta(r) \eta \sigma \tau \omega^x \right] (u,v)^T = 0
\; ,
\ee
it becomes clear the solutions are $\omega^x$ eigenstates,
\be
\left( 
\begin{array}{c}
u \\ v
\end{array}
\right) 
= \left( 
\begin{array}{c}
1 \\ \omega
\end{array}
\right)
 e^{ \eta \sigma \tau \omega \frac{1}{v} \int_0^r dr' \Delta(r') }
\; .
\ee
Already at this point, before taking into account normalizability, 
we see that as many as 16 zero mode solution exist,
parametrized by $\eta,\sigma,\tau,\omega = \pm 1$. Assuming that $\Delta(r)$
is positive at $r \rightarrow \infty$, only half of these 16 solutions are 
exponentially decaying $\eta \sigma \tau \omega = -1$, and thus normalizable 
in an infinite system. The number of zero modes is then reduced to 8. 
In a finite system the exponentially growing solutions correspond to edge states.

The full wavefunction solutions we find are, for $\eta=+1$
\be
\psi_1 = 
\chi_{\tau} \otimes \chi_{\sigma} \otimes
\left( 
\begin{array}{c}
1 \\ 0 \\ \omega \\ 0
\end{array}
\right)
 e^{+ \sigma \tau \omega \frac{1}{v} \int_0^r dr' \Delta(r') }
\; ,
\ee
and for $\eta = -1$
\be
\psi_2 = 
\chi_{\tau} \otimes \chi_{\sigma} \otimes
\frac{1}{r} 
\left( 
\begin{array}{c}
0 \\ e^{i \phi} \\ 0 \\ \omega e^{-i \phi}
\end{array}
\right)
 e^{- \sigma \tau \omega \frac{1}{v} \int_0^r dr' \Delta(r') }
\; .
\ee
Here, as in the previous section, the 4-component vectors correspond to
$(u_1, u_2, v_1,v_2)^T$ ($1,2$ are the sublattice indices). We will follow this convention in the remainder of this section.
With the condition of normalizability, the exponentials must take on the
decaying form $e^{- \frac{1}{v} \int_0^r dr' \Delta(r') }$, and we must have 
$\omega = -\sigma \tau \eta$.
The solutions then become
\be
\psi_{1,2} = 
\chi_{\tau} \otimes \chi_{\sigma} \otimes
\psi_{1,2}^0
 e^{- \frac{1}{v} \int_0^r dr' \Delta(r') }
\; ,
\ee
where
$\psi_1^0 = 
\left( 
1 , 0 , \sigma \tau , 0
\right)^T
$ 
and $
\psi_2^0 =
\frac{1}{r} 
\left( 
0 , e^{i \phi} , 0 , \sigma \tau e^{-i \phi}
\right)^T
$.

Next we turn to normalizability in the $r \rightarrow 0$ limit.
This is determined by whether the integral $\int_0^{r_0} |\psi|^2 r dr$
diverges. The solution $\psi_1$, is clearly normalizable in this region, 
while $\psi_2$ clearly is not. This leaves us with 4 zero modes, rather 
than 8 as we found for the SNS junction, where no analog of the $r \rightarrow 0$
normalizability condition appears. 

Now we turn to the case away from half filling.
The BdG equations in this case read
\be
\left(
\begin{array}{cc}
\mu -i v \left({\vec \eta} \cdot \nabla \right) & \Delta(r) e^{i \phi} \eta^y \sigma \tau \\
\Delta(r) e^{-i \phi} \sigma \tau & -\mu - i v \left({\vec \eta}^* \cdot \nabla \right)
\end{array}
\right)
\psi = 0
\; .
\ee
We can eliminate the $\phi$ dependence from the problem by choosing the 
exact same form as in the half filling case
\be
\psi(r,\phi) = \left(
u_1(r), e^{i \phi } u_2(r),v_1(r),e^{-i \phi } v_2(r)
\right)^T
\; .
\ee
The reduced ODEs then involve only the radial coordinate.

At this point it is useful, to make the educated guess
\be
\psi(r,\phi) = \left(
\begin{array}{c}
u_1(r) \\ 
e^{i \phi } u_2(r) \\
v_1(r) \\
e^{-i \phi } v_2(r)
\end{array}
\right) e^{- \frac{1}{v} \int_0^r dr' \Delta(r')}
\; ,
\ee
inspired by the form of the solution for the SNS junction.
Plugging this form into the ODEs, does not
remove the order parameter from them completely. However, choosing
$v_1(r) = -\sigma \tau u_1(r)$ and $v_2(r) = \sigma \tau u_2(r)$ in addition, does remove
the order parameter. The reduced ODEs, involving only $u_{1,2}$ then read
\be
\begin{split} &
\mu  u_2(r) - i v u_1'(r) = 0 \\ &
\mu  u_1(r) - \frac{i v u_2(r)}{r} - i v u_2'(r) = 0
\; .
\end{split}
\ee
Extracting $u_2$ from  first equation, and plugging it into the second yields a single ODE
for $u_1$
\be
\frac{u_1'(r) v^2}{r \mu }+\frac{u_1''(r) v^2}{\mu }+\mu  u_1(r) = 0
\; .
\ee
The solutions are Bessel functions $J_0(\frac{r \mu}{v}),Y_0(\frac{r \mu}{v})$,
and $u_2$ is obtained from $u_2(r) = \frac{i v u_1'(r)}{\mu }$.
The two zero mode solutions we obtain are then 
\be\label{s_wave_vortex_sol}
\psi = \chi_{\tau} \otimes \chi_{\sigma} \otimes
e^{- \frac{1}{v} \int_0^r dr' \Delta(r')}
\left(
\begin{array}{l}
 J_0\left(\frac{r \mu }{v}\right) \\
 -i e^{i \phi } J_1\left(\frac{r \mu }{v}\right) \\
 -\sigma  \tau  J_0\left(\frac{r \mu }{v}\right) \\
 -i e^{-i \phi } \sigma  \tau  J_1\left(\frac{r \mu }{v}\right)
\end{array}
\right)
\; ,
\ee
and the second solution simply has all the Bessel functions of the 1st kind 
replaced with Bessel functions of the 2nd kind, with the same parameters.
Only the Bessel functions of the 1st kind is normalizable in $r \rightarrow 0$, 
or alternatively (specifically 
$\int_0^{r_0} Y_1\left(\frac{r \mu }{v}\right)^2 r dr$ diverges ), 
if we impose a boundary 
condition at some small $r = a$, we will pick out some combination of the 
two Bessel function kinds.
With the $\sigma,\tau$ degeneracy we end up with 4 zero modes.
It is easy to verify that the BCS particle-hole relation yields 
$\omega^x \left( \psi_{1,2}^{\sigma \tau} \right)^* = - \sigma \tau \psi_{1,2}^{-\sigma,-\tau}$
producing no new zero modes beyond the 4 mandated by the system symmetries.

It is noteworthy that the Bessel function, far from the vortex core $\frac{r \mu }{v} \gg 1$,
has an oscillatory nature, with a length scale $\frac{v}{\mu}$, precisely as in the zero modes 
we find for the SNS and edge states.

\subsection{Singlet p+ip condensate}
\label{p_wave_vortex}

We turn now to the simple $p_x + i p_y$ singlet-pairing condensate phase. 
The vortex core eigenvalue spectrum is determined by the BdG equation
\be\label{BdG4}
\left( {\mathcal H}_0 + {\mathcal H}_2 \right)\psi = E \psi
\; ,
\ee
where \eqref{H_0} and \eqref{H_2} contain the full details of the kinetic and pairing term.

As in the s-wave case, we choose solutions that are $\sigma^y \tau^y$ eigenstates, precisely 
as in the SNS and edge geometries,
and as a result the energy spectrum will be at least 4-fold degenerate.
We will model the vortex by assuming the form $i \Delta_1 = \Delta(r) e^{+i \phi}$, with $\Delta(r)$ real
(different from our conventions in earlier sections so that we can use the same ansatz for the polar angle dependence as for the s-wave case). Also, 
since it will prove convenient, we will assume that the order parameter radial profile is piecewise 
uniform - vanishing inside the vortex core, and constant outside it.

With the insight gained in the previous subsection,
we find the $\phi$ dependence can be eliminated from the zero-mode problem by 
choosing the wavefunction form
\be
\psi(r,\phi) = \left(
u_1(r), e^{i \phi } u_2(r),v_1(r),e^{-i \phi } v_2(r)
\right)^T
\; .
\ee
The reduced ODEs then involve only the radial coordinate, and can be cast
in the form $\partial_r \psi = A \psi$ where
\be
\begin{split}
A =
- \frac{1}{2 r}
- \frac{1}{v^2 + \Delta^2}
\Big[ &
\Delta \mu \sigma \tau \omega^y \eta^z
+ 
i v \mu \eta^x \omega^z
\\ 
+ &
\frac{\Delta}{2 r} v \sigma \tau \omega^x \eta^x
- \eta^z
\frac{v^2}{2 r}
\Big]
\; .
\end{split}
\ee

We first find the asymptotic ($r \rightarrow \infty$) solutions to the ODE system.
Neglecting all the $\frac{1}{r}$ terms, we find
\be
A =
- \frac{1}{v^2 + \Delta^2}
\left[
\Delta \mu \sigma \tau \omega^y \eta^z
+ 
i v \mu \eta^x \omega^z
\right]
\; .
\ee
We can diagonalize the asymptotic form of $A$ with the unitary transformation 
\be
O = \frac{1}{2}
\left(
\begin{array}{cccc}
 i & -i & -i & i \\
 i & i & -i & -i \\
 -1 & 1 & -1 & 1 \\
 1 & 1 & 1 & 1
\end{array}
\right)\; ,
\ee
yielding
\be
\begin{split} &
O^{\dagger} A O = \frac{\mu}{v^2 + \Delta^2} 
\\ &
{\textrm{Diagonal}}(- i v - \Delta \sigma \tau,+ i v - \Delta \sigma \tau,
- i v + \Delta \sigma \tau, + i v + \Delta \sigma \tau
)
\; .
\end{split}
\ee
Using this unitary transformation on the full matrix $A$, we find a block-diagonal form
\be
O^{\dagger} A O = 
\left(
\begin{array}{ll}
{\mathcal A}_{\sigma \tau} & 0 \\ 
0 & {\mathcal A}_{-\sigma \tau}
\end{array}
\right)
\; ,
\ee
where
\be 
\begin{split} &
{\mathcal A}_{\sigma \tau} = - \frac{1}{2 r} 
\\ &
+ \frac{1}{\Delta^2 + v^2}
\left[ 
- \Delta \mu \sigma \tau
+
\left(
\begin{array}{ll}
 -i v \mu  & \frac{i v \Delta  \sigma  \tau }{2
   r}-\frac{v^2}{2 r} \\
 -\frac{v^2}{2 r}-\frac{i \Delta  \sigma  \tau  v}{2 r}
   & i v \mu 
\end{array}
\right)\right]
\; .
\end{split}
\ee
Here the terms outside the matrix are implicitly multiplied by 2 by 2
identity matrices.

We now turn to solve the reduced ODE system
${\mathcal A}_{\sigma \tau} \xi = \partial_r \xi$,
where $\xi = (f_1(r) , f_2(r) )^T$.
The precise solution will prove cumbersome to work with, 
and so we will start with an approximate solution that will 
reveal all the features of the solutions we need to find.
First we write $\xi = \frac{1}{\sqrt{r}} 
e^{- \frac{\Delta \mu \sigma \tau}{\Delta^2 + v^2} r} \zeta$.
We note at this point that for the solution to be normalizable at 
$r \rightarrow \infty$, we must have $\mu \sigma \tau >0$,
if however this is not the case, then we simply choose the 
solution for the ${\mathcal A}_{-\sigma \tau}$ sector, in $O^{\dagger} A O$.
The ODE for $\zeta$ is then
\be
\partial_r \zeta = 
\frac{1}{\Delta^2 + v^2}
\left(
\begin{array}{ll}
 -i v \mu  & \frac{i v \Delta  \sigma  \tau }{2
   r}-\frac{v^2}{2 r} \\
 -\frac{v^2}{2 r}-\frac{i \Delta  \sigma  \tau  v}{2 r}
   & i v \mu 
\end{array}
\right) \zeta
\; .
\ee
Now we assume that $v \gg \Delta$, and consider the small $r$ limit, so that we can
approximate
\be
\partial_r \zeta \approx
- \frac{1}{2 r}
\left(
\begin{array}{ll}
 0 & 1 \\
 1 & 0
\end{array}
\right) \zeta
\; ,
\ee
for which the solutions are 
\be
\zeta_{\eta} = (1, \eta)^T r^{-\eta/2} 
\; .
\ee
In the limit $r \rightarrow 0$, only the $\eta = +1$ solution is normalizable.
We therefore find approximate solutions, that have only the $\sigma, \tau$ 
4-fold degeneracy.

We now turn to briefly make connection with the precise solutions for the zero modes.
From the equation $\partial_r f_1 = \ldots$, we extract $f_2(r)$,
and substitute it in the other equation. This yields a single 2nd order ODE
\be
\begin{split} &
\left(\Delta^2 + 8 r \mu \sigma \tau \Delta + 4 r \mu (i v + r \mu ) \right) f_1(r)
\\ & + 
8 r \left(v^2 + \Delta^2 + r \Delta \mu \sigma \tau \right) f_1'(r)
\\ & +
4 r^2 \left(v^2+\Delta ^2\right) f_1''(r) = 0
\; .
\end{split}
\ee
Next, we write 
$f_1(r) = g(r) e^{-\frac{r \mu  (i v+\Delta  \sigma  \tau )}{v^2+\Delta ^2}} r^{\frac{1}{2} \left(\frac{v}{\sqrt{v^2+\Delta ^2}}-1\right)}$,
and furthermore replace the radial variable with $r = -\frac{i z \left(v^2+\Delta ^2\right)}{2 v \mu }$.
The ODE for $g(r)$ is then 
\be
-\frac{v g(z)}{2 \sqrt{v^2+\Delta ^2}}+\left(\frac{v}{\sqrt{v^2+\Delta ^2}}-z+1\right) g'(z)+z g''(z) = 0
\; ,
\ee
which is the confluent hypergeometric ODE.
The solutions are
\be
g(z) = c_1 M(a,1+2a,z) + c_2 z^{-2 a} M(-a,1-2a,z)
\; ,
\ee
where $c_{1,2}$ are free coefficients, $a = \frac{v}{2 \sqrt{v^2 + \Delta^2}} \leq \frac{1}{2}$
(and also $a > 0$),
and $M(a,b,z) = \, _{1} \! F_1(a,b,z)$ is the confluent hypergeometric function of the first kind 
(or Kummer function). With the complex valued variable $z \sim i r$, the solutions are 
well-behaved at $r \rightarrow \infty$. At small $r$, to lowest order 
$M(a,b,z) \approx 1 + O(z)$, and the radial part of the wavefunction behaves like
$f_1(r) \sim r^{-1/2} r^a g(z) \sim 
r^{-1/2} r^a \left( c_1 + c_2 r^{-2 a} + \ldots \right) 
\sim c_1 r^{a-1/2} + c_2 r^{-a-1/2} \sim c_2 r^{-a-1/2}$.
When $a<\frac{1}{2}$ this solution is normalizable. However, in the $r \rightarrow 0$
limit the order parameter must vanish, so we must take $\Delta = 0$, in which case $a=1/2$,
and causes a logarithmic divergence when we try to normalize it.
Therefore, including the $r \rightarrow 0$ point, we must set $c_2 = 0$,
and we are indeed left with only 4 solutions for zero modes.

\subsection{Quadratic correction for the vortex core zero modes}
\label{quad_correction_s_wave}

In both cases of the s-wave and $p_x + i p_y$ spin singlet phases, the eigenstates are angular 
momentum eigenstates as well, and have the general separable wavefunction form
\be
\psi_{\ell}(r,\phi) = 
\chi_{\sigma} \otimes \chi_{\tau} \otimes
e^{i \ell \phi} 
\left(
\begin{array}{c}
u_1(r) \\
e^{i \phi} u_2(r) \\
v_1(r) \\
e^{- i \phi} v_2(r) 
\end{array}
\right)
\; .
\ee
We take the quadratic correction \eqref{H_4} in polar coordinates
and find that it takes this wavefunction into the form
\be
{\mathcal H}_4 \psi_{\ell}(r,\phi) = 
\chi_{\sigma} \otimes \chi_{-\tau} \otimes
e^{i \ell \phi} 
\left(
\begin{array}{c}
e^{3 i \phi} f_1(r) \\
e^{-2 i \phi} f_2(r) \\
e^{-3  i \phi} g_1(r) \\
e^{2 i \phi} g_2(r) 
\end{array}
\right)
\; .
\ee
Then trying to take the product $\bra{\psi_{\ell'}} {\mathcal H}_4 \ket{\psi_{\ell}}$,
it suffices to consider the angular dependency
\be
\begin{split} &
\bra{\psi_{\ell'}} {\mathcal H}_4 \ket{\psi_{\ell}} 
\\ &
\sim
\int_0^{2 \pi} d\phi
e^{i (\ell' - \ell) \phi}
\left(
u_1^* ,
e^{-i \phi} u_2^* ,
v_1^* ,
e^{+i \phi} v_2^* 
\right)
\cdot
\left(
\begin{array}{c}
e^{3 i \phi} f_1(r) \\
e^{-2 i \phi} f_2(r) \\
e^{-3  i \phi} g_1(r) \\
e^{2 i \phi} g_2(r)
\end{array}
\right)
\\ &
\sim
\int_0^{2 \pi} d\phi
e^{i (\ell' - \ell) \phi}
\Big[ 
u_1^* f_1 e^{3 i \phi} + u_2^* f_2 e^{-3 i \phi} 
\\ & 
+ v_1^* g_1 e^{-3 i \phi} + v_2^* g_2 e^{3 i \phi}
\Big]
\; .
\end{split}
\ee
from the polar phase integration we conclude that
nonzero matrix elements exist only between states with angular momentum $\ell$ differing by $\pm 3$.
In particular, to first order, there is no correction, and the zero modes persist to this order. 

To conclude this section, we point out that the pure Dirac theory
approximating the honeycomb lattice allows
for topological zero modes to appear bound to vortices in both spinfull 
condensate phases.

\section{Numerics on the honeycomb lattice model}
\label{Numerics}

In the previous sections we found zero modes exist in both the s-wave and $p_x + i p_y$
spin-singlet states in the continuum Dirac approximation for the honeycomb lattice.
We also tried to ascertain whether the zero modes
exist beyond the approximate Dirac theory for the honeycomb lattice,
by taking into account the quadratic correction to the kinetic energy Eq.~\eqref{H_4}.
We calculated whether this correction splits the zero modes at first order in perturbation theory.
With the exception of one case, we always found that to first order, no splitting occurs.
The exception is the s-wave phase in the SNS geometry with $\alpha \neq 0$ (Section \ref{s_wave_SNS}), where we found the correction 
\emph{does} give a splitting to first order in the quadratic correction.
In contrast, in the edge state and vortex cases, no such splitting occurred at first order.
While the SNS splitting does vanish when the junction boundaries are of the armchair edge type,
consistent with the edge state result, the collection of these results is inconclusive
as to whether the zero modes really do appear in the lattice model, and not just in the idealized 
approximate Dirac theory. To answer this question definitively, we have performed numerical 
calculations (exact diagonalization) on the precise lattice models for all phases where we suspect zero modes occur.




\begin{figure}
	\centering
		\includegraphics[width=3.0in]{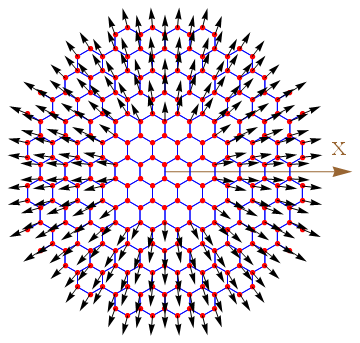}
	\caption{(color online) Circular lattice patch in the vortex state. The (black) arrows represent the phase 
	at a lattice point as the angle the arrow makes with the x-axis in the picture. The (red) dots 
	are the honeycomb lattice sites, and (blue) lines are the nearest neighbor links. Here, for illustration 
	purposes, the radius of the vortex core is taken to be $2$ (the radius of the lattice patch is $6$, 
	where the nearest neighbor distance is $1/\sqrt{3}$).
	Following the arrows it can be verified that the vortex indeed has a unit vorticity.}
	\label{fig:Generic_numerical_BdG_honeycomb_circle}
\end{figure}

\begin{figure}
     \centering
     \subfigure[]{
          \label{fig:square_scaling}
          \includegraphics[height=1.2in]{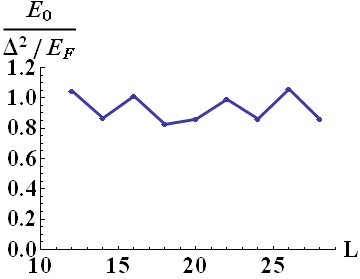}}
     \subfigure[]{
          \label{fig:circular_scaling}
          \includegraphics[height=1.2in]{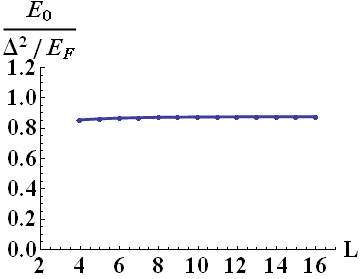}}\\
     \subfigure[]{
           \label{fig:rectangular_scaling}
           \includegraphics[height=1.2in]{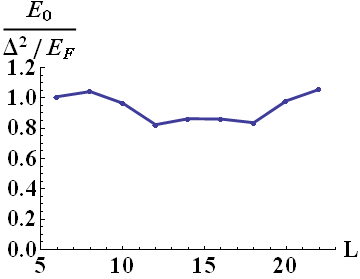}}
     \subfigure[]{
           \label{fig:elliptic_scaling}
          \includegraphics[height=1.2in]{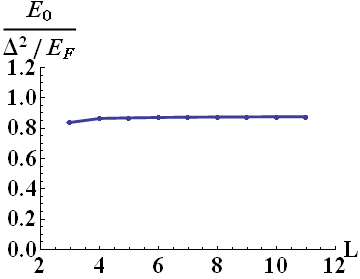}}
     \caption{Scaling of the lowest energy with lattice patch size. 
	We plot the ratio of the lowest energy to the de Gennes energy for all sizes we take. 
	The square lattice patch\subref{fig:square_scaling} is of size $L \times L$, 
	the circular lattice patch\subref{fig:circular_scaling} is of radius $L$,
	the rectangular lattice patch\subref{fig:rectangular_scaling} is of dimensions $L \times \frac{3}{2} L$, and
	the elliptic lattice patch\subref{fig:elliptic_scaling} is of main axis' $L$ and $2 L$.
	The corresponding number of sites for each lattice patch is detailed in table~\ref{tab:LatticePatchSizes}.
	It is clear from the graphs that the lowest energy is always of the order of the de Gennes energy, and remains
	roughly unchanged when we increase the system size.}
     \label{fig:energy_scaling}
\end{figure}

\begin{figure}
     \centering
     \subfigure[]{
          \label{fig:square_p_wave_scaling}
          \includegraphics[height=1.2in]{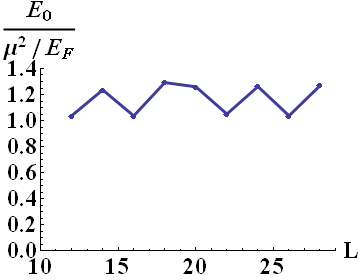}}
     \subfigure[]{
          \label{fig:circular_p_wave_scaling}
          \includegraphics[height=1.2in]{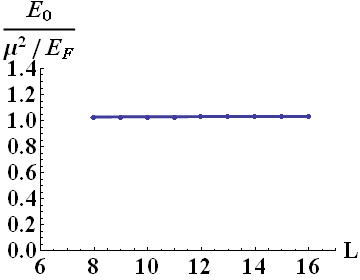}}\\
     \subfigure[]{
           \label{fig:rectangular_p_wave_scaling}
           \includegraphics[height=1.2in]{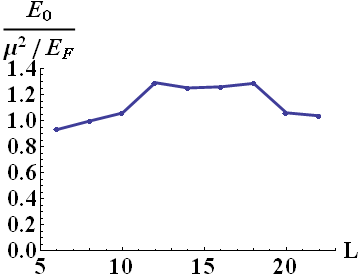}}
     \subfigure[]{
           \label{fig:elliptic_p_wave_scaling}
          \includegraphics[height=1.2in]{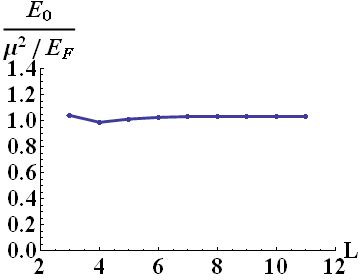}}
     \caption{Scaling of the lowest energy with lattice patch size. 
	We plot the ratio of the lowest energy to the de Gennes energy for all sizes we take,
	for the spin-singlet $p_x + i p_y$ state. In this case, the gap energy scales like the chemical potential $\mu$,
	so we take as the de Gennes energy scale $\mu^2/E_F$.
	The square lattice patch\subref{fig:square_p_wave_scaling} is of size $L \times L$, 
	the circular lattice patch\subref{fig:circular_p_wave_scaling} is of radius $L$,
	the rectangular lattice patch\subref{fig:rectangular_p_wave_scaling} is of dimensions $L \times \frac{3}{2} L$, and
	the elliptic lattice patch\subref{fig:elliptic_p_wave_scaling} is of main axis' $L$ and $2 L$.
	The corresponding number of sites for each lattice patch is detailed in table~\ref{tab:LatticePatchSizes}.
	It is clear from the graphs that the lowest energy is always of the order of the de Gennes energy, and remains
	roughly unchanged when we increase the system size.}
     \label{fig:energy_p_wave_scaling}
\end{figure}

\begin{figure}
     \centering
     \subfigure[]{
          \label{fig:square_low_s_wave}
          \includegraphics[height=1.5in]{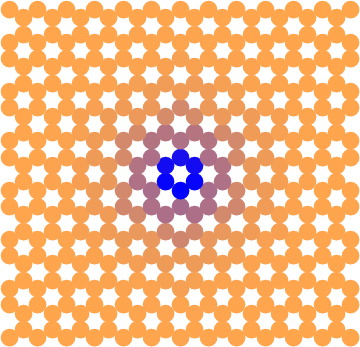}}
     \subfigure[]{
          \label{fig:circular_low_s_wave}
          \includegraphics[height=1.5in]{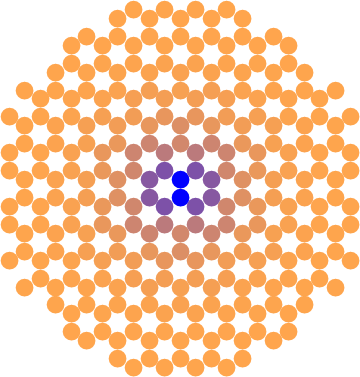}}\\
     \subfigure[]{
           \label{fig:rectangular_low_s_wave}
           \includegraphics[height=2.0in]{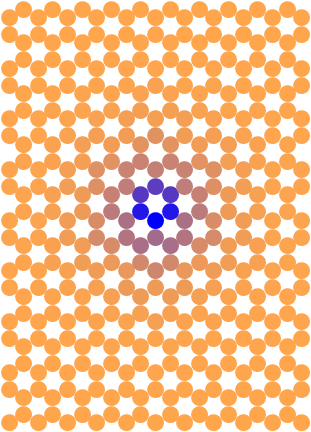}}
           \hspace{0.3in}
     \subfigure[]{
           \label{fig:elliptic_low_s_wave}
          \includegraphics[height=2.0in]{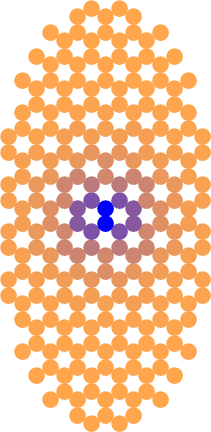}}
     \caption{The lowest energy quasiparticle state wavefunction density in the s-wave state,
     in the square\subref{fig:square_low_s_wave}, 
     circular\subref{fig:circular_low_s_wave}, 
     rectangular\subref{fig:rectangular_low_s_wave} 
     and elliptic\subref{fig:elliptic_low_s_wave} geometries.
     The density $|u_j|^2 + |v_j|^2$ at each site $j$ is represented
     by the color of the dot at each lattice site. 
     The highest density is colored blue (black), and zero density is colored orange(light gray).
     It is clear in all cases that the lowest energy state is bound to vortex core at center of the geometry.
     Here the vortex core size was taken to be $0$. With the nearest neighbor distance taken to be $1/\sqrt{3}$,
     the square\subref{fig:square_low_s_wave} lattice patch has dimensions $12 \times 12$ ($350$ sites),
     the circular\subref{fig:circular_low_s_wave} lattice patch has radius of $6$ ($262$ sites),
     the rectangular\subref{fig:rectangular_low_s_wave} lattice patch has dimensions $10 \times 15$ ($357$ sites),
     and the elliptic\subref{fig:elliptic_low_s_wave} lattice patch has main axis of length $4$ and $8$ ($222$ sites).
     }
     \label{fig:s_wave}
\end{figure}

We consider the vortex state case 
of the two spin-singlet phases at precisely half filling. First, we constructed lattice patches of square, rectangular, circular and elliptic 
shapes (see Fig.~\ref{fig:s_wave} for illustrations), of various sizes. 
We then diagonalized the matrices describing the 
lattice model \eqref{lattice_model} on these lattice patches, with the 2 spin-singlet order parameters 
including unit-vorticity vortices at their centers (one representative example is shown 
in Fig.\ref{fig:Generic_numerical_BdG_honeycomb_circle}). 
In all cases, we find the lowest
energy eigenvalues $E_0$, and compare them with the de Gennes energy scale $\frac{E_g^2}{E_F}$,
the energy scale one expects for vortex core bound state\cite{DeGennes:64}. 
For the s-wave state, the gap energy $E_g = \Delta$, while for the spin-singlet $p_x + i p_y$
state, the gap scales with the chemical potential~\cite{Uchoa:prl07} $E_g \sim \mu$.
The sizes of the various lattice patch geometries we take is detailed in table~\ref{tab:LatticePatchSizes},
where the nearest neighbor distance is $1/\sqrt{3}$ (the primitive Bravais lattice vectors are then of length $1$). 
The scaling of the lowest energy with the finite system size is described in 
Fig.~\ref{fig:energy_scaling} for the s-wave state, and in Fig.~\ref{fig:energy_p_wave_scaling} for
the spin-singlet $p_x + i p_y$ state. It is clear from all scaling plots that the lowest energy is of order of the 
de Gennes energy, and that this energy does not significantly decrease with growing system size.
This would indicate that there exist no zero modes in these phases, despite the results from the 
continuum Dirac theory.

\begin{table}
	\centering
		\begin{tabular}{| l || c | l || c | l || c | l || c |}
		\hline
			\multicolumn{2}{|c|}{Square} & \multicolumn{2}{c|}{Circular} & 
			\multicolumn{2}{c|}{Rectangular} & \multicolumn{2}{c|}{Elliptic} \\ 
			\hline
			$L$ & sites & $L$ & sites & $L$ & sites & $L$ & sites \\
			\hline
			12 & 350 & 8 & 466 & 6 & 137 & 3 & 130 \\
			14 & 464 & 9 & 590 & 8 & 238 & 4 & 222 \\
			16 & 611 & 10 & 726 & 10 & 357 & 5 & 358 \\
			18 & 777 & 11 & 874 & 12 & 525 & 6 & 512 \\
			20 & 943 & 12 & 1036 & 14 & 696 & 7 & 700 \\
			22 & 1147 & 13 & 1226 & 16 & 924 & 8 & 922 \\
			24 & 1372 & 14 & 1416 & 18 & 1147 & 9 & 1162 \\
			26 & 1590 & 15 & 1630 & 20 & 1415 & 10 & 1446 \\
			28 & 1824 &  16 & 1858 & 22 & 1710 & 11 & 1746 \\
			\hline
		\end{tabular}
	\caption{Lattice patch sizes}
	\label{tab:LatticePatchSizes}
\end{table}

\begin{figure}
     \centering
     \subfigure[]{
          \label{fig:square_low_p_wave}
          \includegraphics[height=1.5in]{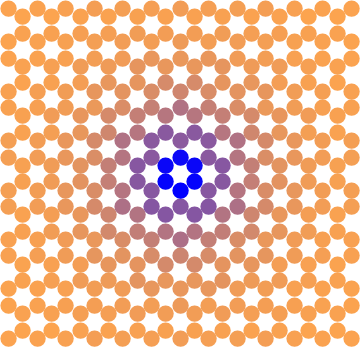}}
     \subfigure[]{
          \label{fig:circular_low_p_wave}
          \includegraphics[height=1.5in]{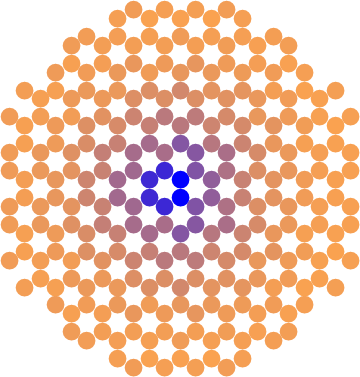}}\\
     \subfigure[]{
           \label{fig:rectangular_low_p_wave}
           \includegraphics[height=2.0in]{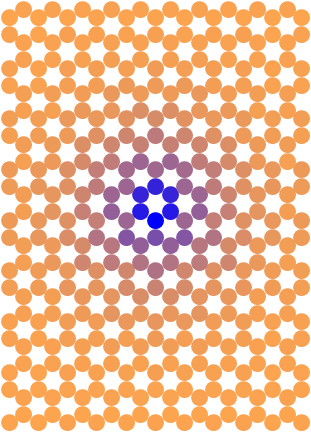}}
           \hspace{0.3in}
     \subfigure[]{
           \label{fig:elliptic_low_p_wave}
          \includegraphics[height=2.0in]{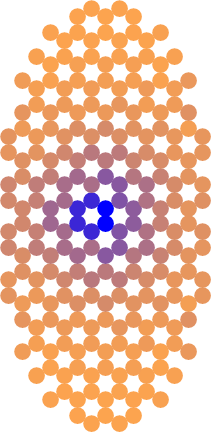}}
     \caption{The lowest energy quasiparticle state wavefunction density in the spin-singlet $p_x + i p_y$ state,
     in the square\subref{fig:square_low_p_wave}, 
     circular\subref{fig:circular_low_p_wave}, 
     rectangular\subref{fig:rectangular_low_p_wave} 
     and elliptic\subref{fig:elliptic_low_p_wave} geometries.
     The density $|u_j|^2 + |v_j|^2$ at each site $j$ is represented
     by the color of the dot at each lattice site. 
     The highest density is colored blue (black), and zero density is colored orange(light gray).
     It is clear in all cases that the lowest energy state is bound to vortex core at center of the geometry.
     Here the vortex core size was taken to be $0$. With the nearest neighbor distance taken to be $1/\sqrt{3}$,
     the square\subref{fig:square_low_s_wave} lattice patch has dimensions $12 \times 12$ ($350$ sites),
     the circular\subref{fig:circular_low_s_wave} lattice patch has radius of $6$ ($262$ sites),
     the rectangular\subref{fig:rectangular_low_s_wave} lattice patch has dimensions $10 \times 15$ ($357$ sites),
     and the elliptic\subref{fig:elliptic_low_s_wave} lattice patch has main axis of length $4$ and $8$ ($222$ sites).
     }
     \label{fig:p_wave}
\end{figure}

In addition, we plot the spatial density of the lowest energy quasiparticle density on the lattice patch,
in a number of representative cases in Fig.~\ref{fig:s_wave} and Fig.~\ref{fig:p_wave},
in order to verify that these indeed are vortex core bound states.
Specifically, at each lattice site $j$ we plot a dot with its color signifying the relative
magnitude of $|u_j|^2 + |v_j|^2$ (the values are normalized to run between 0 and 1, and the color is 
varied linearly with with this value). From the plots it is clear that these lowest energy states are 
indeed vortex core bound states.

\begin{figure}
	\centering
		\includegraphics[width=3.0in]{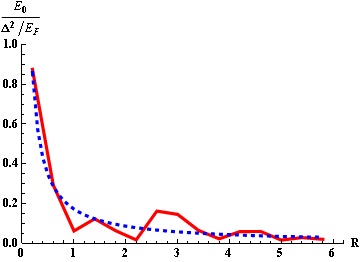}
	\caption{Ratio of lowest energy scale to de Gennes energy scale, with varying vortex core size.
	The vortex core sizes range from $0.2$ to $6.0$ in increments of $0.4$, and the lattice patch 
	is circular with a radius of $6.0$. The nearest neighbor distance is $1/\sqrt{3}$. The raw data 
	is denoted by the continuous (red) curve, and the fit is denoted by a dashed (blue) curve.}
	\label{fig:vortex_core_size_fit}
\end{figure}

In all the cases described in Fig.~\ref{fig:energy_scaling} and Fig.~\ref{fig:energy_p_wave_scaling}, the 
vortex core size was taken to be $0$, forcing the 
vortex phase winding to occur over a distance that is comparable to the lattice length scale.
This fact is what invalidates the Dirac continuum theory - the order parameter in these cases is \emph{not}
a slowly varying function on the lattice scale, near the vortex core.
Following this last observation, we also calculated the energy spectrum for a series of different 
vortex core sizes, ranging from $0.2$ to $5.8$ in increments of $0.4$, 
while keeping fixed the overall system size (circular lattice patch of radius $12.0$). 
The results are plotted in Fig.~\ref{fig:vortex_core_size_fit} and it is clear from them that the energy splitting decreases rapidly 
with the vortex core size. The highest energy we find (at the smallest radius) is $0.88$ times the de 
Gennes scale, and the smallest energy scale we find is $0.0016$ (at a radius of $5.0$).
However, the de Gennes energy scale in terms of the correlation length of a superconductor is actually 
$\frac{\Delta}{k_F \xi} $\cite{DeGennes:64}. From this we expect that the lowest energy should change with the 
vortex core radius $R$ as $E_0 \sim \frac{1}{R}$. We fit the raw data in Fig.~\ref{fig:vortex_core_size_fit} to 
a curve $-0.0007 + 0.17/R$, also displayed in Fig.~\ref{fig:vortex_core_size_fit}.
When the vortex core size is bigger, the phase winding occurs over a larger distance, and the
approximation of a slowly varying order parameter improves, but the lowest energy is \emph{still} 
of the de Gennes scale. We find therefore that the zero modes are split, and correspond to the de Gennes
bound state spectrum.

For a vortex one would expect the core size to be of the order of the correlation length in the superconducting 
state. The parameters we use are $\Delta = 0.5, t = 1.0$ the lattice constant $a = 1.0$ and working in units where $\hbar = 1$, we have a correlation length that is of the order of the lattice constant
$\xi \sim \frac{\hbar v_F}{\pi \Delta} \sim \frac{\hbar 3 t a}{2 \pi \Delta} \sim 1$.
In Fig.~\ref{fig:vortex_core_size_fit} we can see that for this core size, the lowest energy is
between $5$ and $10$ times smaller than the de Gennes scale. 

In conclusion, the numerics we have done show that the zero modes appearing in the continuum Dirac theory
are split in the lattice model, but that the bound state energy can be significantly smaller than the 
de Gennes scale.

\section{Experimental realizations - Bose Fermi mixtures}

\label{ExperimentalRealizations}

Superconductivity, which appears in many conventional fermionic systems, does not seem to occur 
intrinsically in graphene, the most readily available realization of the honeycomb tight binding model, 
but can be induced via the proximity effect close to another 
superconducting material\cite{Shailos:2007,heersche-2007-446}. 
The peculiar band structure of the honeycomb lattice is not however limited to graphene - it is 
just one material realization (other possibilities may include thin films of quasi-2D honeycomb layered superconductors\cite{PhysRevLett.14.225,Kempa:2000,Kopelevich:2000}). 
Another possibility is a cold fermion gas trapped in an optical lattice,
with a fermionic atom density corresponding to about half filling. Since the atoms are electrically neutral,
the interactions between them are to a good approximation simple collisions, corresponding to an 
on-site interaction in a lattice model. Superconductivity requires some attraction between fermions, so
in order to have any hope of realizing such a phase, one needs to cause attractive interactions between
the fermions. In solids, phonons provide this mechanism by inducing an attractive interaction between electrons. In cold atom gases, this role can be assumed by adding bosonic atoms; sound modes in a Bose-Einstein condensate mimic phonons in a solid. The virtues of the cold atom realization do not end in simply making the superconducting state feasible, but also provide a great deal of control over many parameters.

Motivated by the reasoning discussed in the previous paragraph, we will now consider a model of a 
Bose-Fermi mixture on the honeycomb lattice model. We consider only on-site interactions, since
usually one has to work quite hard to make longer range interactions appreciable compared 
to them in cold atom systems.
Our Hamiltonian therefore reads
\be\label{Bose_Fermi}
\begin{split}
{\mathcal H} = & 
- t \sum_{\langle i j\rangle, \alpha} f_{i \alpha}^{\dagger} f_{j \alpha} 
+ \mu \sum_{j ,\alpha} f_{j \alpha}^{\dagger} f_{j \alpha} 
\\ &
- w \sum_{\langle i j\rangle} b_i^{\dagger} b_j + 3 w \sum_j b_j^{\dagger} b_j
\\ &
+ U_{bb} \sum_j \left[ \left( b_j^{\dagger} b_j \right) - \rho_0 \right]^2 
+ U_{ff} \sum_j \left( f_{j \uparrow}^{\dagger} f_{j \uparrow} \right) 
								\left( f_{j \downarrow}^{\dagger} f_{j \downarrow} \right)
\\ &
+ U_{bf} \sum_j \sum_{\alpha} \left( f_{j \alpha}^{\dagger} f_{j \alpha} \right) \left( b_j^{\dagger} b_j \right)
\; ,
\end{split}
\ee
where $b_j$ are bosonic operators and $f_{j,\alpha}$ are the fermionic operators used 
in our manuscript.
As before $i,j$ are used to denote lattice sites, and the 2 greek letters
$\alpha, \beta$ will be used to denote the spin indices $\uparrow, \downarrow$.
The bosonic chemical potential is tuned to the value $3 w$ - so that the bosonic 
band minimum is at zero energy, and $\rho_0$ is the boson density per site. 

With no interaction between the boson and fermions, for small $U_{\textrm{bb}}/w \ll 1$
the bosons will condense into a superfluid state. We assume we are deep in such a phase, 
and that the interactions with the fermions do not destroy the boson superfluidity.
We then use standard Bogoliubov theory to approximate the momentum space 
bose operators $ b_{\mu}({\bf q}) \approx \sqrt{N_0} \delta({\bf q}) + b_{\mu}({\bf q}) $
where the second term is implicitly taken for only nonzero momentum.
Here $N_0$ is half the total number of bosons. The bosonic density operator then becomes
$
\rho_{\mu}({\bf q}) \equiv
\sum_{\bf k} b_{\mu}^{\dagger}({\bf k-q}) b_{\mu}^{\phantom\dagger}({\bf k})
\approx N_0 \delta({\bf q}) + \sqrt{N_0} \left[ 
b_{\mu}^{\dagger}(-{\bf q})  + b_{\mu}^{\phantom\dagger}({\bf q})
\right]
$.
Using the Bogoliubov approximation we expand the bosonic terms of \eqref{Bose_Fermi}
to quadratic order in the operators $b_{\mu}({\bf q})$. Our goal is then to integrate out the 
bosonic degrees of freedom (the action is now Gaussian in the bosonic fields), and in this 
way find the effective Fermionic interactions that are induced.

Taking only those terms in \eqref{Bose_Fermi} involving bosonic operators, and performing a Fourier transformation
we find
\be
\begin{split}
{\mathcal H}_b = & w
\sum_{\bf q, \mu \nu}
b_{\mu}^{\dagger}({\bf q}) \left(3 \delta_{\mu \nu} -\Gamma_{\mu \nu}({\bf q}) \right)
b_{\nu}^{\phantom\dagger}({\bf q})
\\
+ & U_{bf}/N \sum_{\bf q} \sum_{\mu} F_{\mu}({\bf q}) \rho_{\mu}(-{\bf q})
\\
+ & U_{bb}/N \sum_{\bf q} \sum_{\mu} \left[\rho_{\mu}({\bf q}) - N \rho_0 \delta({\bf q}) \right] 
                                     \left[ \rho_{\mu}(-{\bf q}) - N \rho_0 \delta({\bf q}) \right]
\end{split}
\ee
where we have introduced the fermionic density operator
$
F_{\mu}({\bf q}) = 
\sum_{{\bf k}, \alpha} f_{\mu \alpha}^{\dagger}({\bf k-q}) f_{\mu \alpha}^{\phantom\dagger}({\bf k})
$.
It is worth mentioning at this point that the boson density per site $\rho_0 = N_0/N$
where $N$ is the number of unit cells of the lattice.

Assuming $N_0$ is a macroscopic number, we can expand ${\mathcal H}_b$
in powers of $N_0$. Keeping only the leading terms, we are left with a quadratic form in the bosonic operators.
Next we apply a unitary transformation that diagonalizes the hopping term
$b_1({\bf q}) = \frac{1}{\sqrt{2}} e^{+i\theta/2} \left[ a_1({\bf q}) + a_2({\bf q}) \right] $
and
$b_2({\bf q}) = \frac{1}{\sqrt{2}} e^{-i\theta/2} \left[ a_1({\bf q}) - a_2({\bf q}) \right] $
where $e^{+i\theta} = \gamma({\bf q})/|\gamma({\bf q})|$ (which implicitly
depends on ${\bf q}$). At this point it is worthwhile mentioning that 
$\theta(-{\bf q}) = -\theta({\bf q})$, which is extremely useful in the detailed steps of our calculation 
that have been omitted here.
After some rewriting of the Hamiltonian, we arrive at the remarkably separable form
\be
\begin{split}
{\mathcal H}_b \approx &
\sum_{\bf q} \sum_{\mu} \left( g + \epsilon_{\mu}({\bf q}) \right) a_{\mu}^{\dagger}({\bf q}) a_{\mu}^{\phantom\dagger}({\bf q})
\\
+ & \frac{g}{2} \sum_{{\bf q}, \mu} 
\left[
a_{\mu}({\bf q}) a_{\mu}(-{\bf q}) + h.c.
\right]
\\
+ & \frac{U_{bf} \sqrt{N_0}}{N}  \sum_{\bf q}
\Big[ \frac{1}{\sqrt{2}}
\left(
F_1^{\dagger}({\bf q}) e^{+i\theta/2} + F_2^{\dagger}({\bf q}) e^{-i\theta/2}
\right) a_1^{\phantom\dagger}({\bf q})
\\ + &
\frac{1}{\sqrt{2}}
\left(
F_1^{\dagger}({\bf q}) e^{+i\theta/2} - F_2^{\dagger}({\bf q}) e^{-i\theta/2}
\right) a_2^{\phantom\dagger}({\bf q})
+ h.c. \Big] + \ldots
\; ,
\end{split}
\ee
where we have introduced the coupling $g = 2 U_{bb} \rho_0$,
and the band dispersions $\epsilon_1 = w \left( 3 - |\gamma({\bf q})| \right)$,
$\epsilon_2 = w \left( 3 + |\gamma({\bf q})| \right)$.
Note that all the momentum summations above formally exclude the ${\bf q} = 0$
mode. This will hold throughout the remainder of this section, and so it will remain implicit.

Next we employ a Bogoliubov transformation for each of the two bands ($a_{1,2}$)
separately. 
This is accomplished by the transformation
$
a_{\mu}({\bf q}) = u_{\mu}({\bf q}) B({\bf q}) + v_{\mu}({\bf q}) B^{\dagger}(-{\bf q})
$
with
$
u_{\mu}({\bf q}) = \sqrt{\frac{E_{\mu}({\bf q}) + \epsilon_{\mu}({\bf q}) + g}{2 E_{\mu}({\bf q})}}
$ and
$
v_{\mu}({\bf q}) = -\sqrt{\frac{-E_{\mu}({\bf q}) + \epsilon_{\mu}({\bf q}) + g}{2 E_{\mu}({\bf q})}}$
where $E_{\mu}({\bf q}) = \sqrt{\epsilon_{\mu}({\bf q})\left( \epsilon_{\mu}({\bf q}) + 2g \right)}$.
The new operators $B({\bf q})$ are canonical bosons, and
the Hamiltonian takes the form
\be
\begin{split}
{\mathcal H}_b \approx &
\sum_{\bf q} \sum_{\mu} E_{\mu} B_{\mu}^{\dagger} B_{\mu}^{\phantom\dagger}
\\
+ & \frac{U_{bf} \sqrt{N_0}}{\sqrt{2} N} \sum_{\bf q}
\Big[
\left(
F_1^{\dagger} e^{+i\theta/2} + F_2^{\dagger} e^{-i\theta/2}
\right) 
\left( u_1 + v_1 \right)
B_1^{\phantom\dagger}
\\ + &
\left(
F_1^{\dagger} e^{+i\theta/2} - F_2^{\dagger} e^{-i\theta/2}
\right) 
\left( u_2 + v_2 \right)
B_2^{\phantom\dagger}+ h.c. \Big] + \ldots
\; ,
\end{split}
\ee
where for the sake of brevity, we have omitted explicit mention of the ${\bf q}$
dependence of all the operators and functions such as $v,u,E$.

In order to integrate out the bosonic fields, we must pass to a path integral formalism, taking 
into account the imaginary time derivative. Rewriting this term using the bosonic operators (or complex
fields) $B_{\mu}({\bf q})$ we find the action
\be
\begin{split}
{\mathcal S} = &
\int_0^{1/T} \!\!\!\!\! d\tau \sum_j b_j^{\dagger}(\tau) \partial_{\tau} b_j^{\phantom\dagger}(\tau)
- \int_0^{1/T} \!\!\!\!\! d\tau {\mathcal H}_b 
\\ &
= 
\int_0^{1/T} \!\!\!\!\! d\tau \sum_q \sum_{\mu} 
B^{\dagger} \partial_{\tau} B^{\phantom\dagger}
- \int_0^{1/T} \!\!\!\!\! d\tau {\mathcal H}_b
\; ,
\end{split}
\ee
where for brevity we have written $B^{\phantom\dagger} \equiv B_{\mu}^{\phantom\dagger}({\bf q},\tau)$.
Integrating out the bosonic fields, we find
\be
S_{\textrm{eff}} = - \sum_{{\bf q},\mu, \omega_n} h_{\mu}^{\dagger}({\bf q},\omega_n)
h_{\mu}^{\phantom\dagger}({\bf q},\omega_n) 
\frac{1}{E_{\mu}({\bf q}) - i \omega_n}
\ee
where
\be
\begin{split} &
h_1^{\dagger} = 
\frac{U_{bf} \sqrt{N_0}}{\sqrt{2} N}
\left(
F_1^{\dagger} e^{+i\theta/2} + F_2^{\dagger} e^{-i\theta/2}
\right) 
\left( u_1 + v_1 \right)
\\ &
h_2^{\dagger} = 
\frac{U_{bf} \sqrt{N_0}}{\sqrt{2} N}
\left(
F_1^{\dagger} e^{+i\theta/2} - F_2^{\dagger} e^{-i\theta/2}
\right) 
\left( u_2 + v_2 \right)
\; .
\end{split}
\ee

If we consider only low frequency effective interactions ($\omega_n \rightarrow 0$),
then we can return to a Hamiltonian formulation of our problem with
\be\label{Eff_Interaction}
\begin{split}
{\mathcal H}_{\textrm{eff}} & = - \sum_{{\bf q},\mu} h_{\mu}^{\dagger}({\bf q})
h_{\mu}^{\phantom\dagger}({\bf q}) 
\frac{1}{E_{\mu}({\bf q})}
\\ & =
- \frac{1}{N}
\sum_{{\bf q}, \mu, \nu}
F_{\mu}^{\dagger}({\bf q}) V_{\mu \nu}({\bf q}) F_{\nu}({\bf q})
\; .
\end{split}
\ee
The interaction vertex we have introduced $V_{\mu \nu}({\bf q})$
has components
\be
\begin{split}
V_{1 1}({\bf q}) = & V_{2 2}({\bf q}) = U_{b f}^2 \frac{\rho_0}{2} 
\left[ \frac{(u_1+v_1)^2}{E_1} + \frac{(u_2+v_2)^2}{E_2} \right]
\\
V_{1 2}({\bf q}) = & V_{2 1}(-{\bf q}) = V_{2 1}^*({\bf q}) 
\\ = &  
e^{+i \theta}
U_{b f}^2 \frac{\rho_0}{2} 
\left[ \frac{(u_1+v_1)^2}{E_1} - \frac{(u_2+v_2)^2}{E_2} \right]
\end{split}
\; .
\ee
Note that $V_{i j}$ will be invariant under 
all the symmetries of the honeycomb lattice, including all lattice translations.

Assuming the fermi level passes near the Dirac nodes of the honeycomb band structure (near half filling).
The most significant low-energy excitations of the system then involve the fermionic 
operators with momenta near the Dirac nodes, at $\pm {\bf Q}$.
The density fluctuation operator
$
F_{\mu}({\bf q}) = 
\sum_{k, \alpha} f_{\mu \alpha}^{\dagger}({\bf k} - {\bf q}) f_{\mu \alpha}^{\phantom\dagger}({\bf k})
$
will be significant only when ${\bf k} = \pm {\bf Q}$ and ${\bf k} - {\bf q} = \pm {\bf Q}$.
This results in 3 possible regions for the exchange momentum: \newline
(i) ${\bf q} \approx 0$ \newline
(ii) ${\bf q} \approx 2 {\bf Q} = - {\bf Q}$ \newline
(iii) ${\bf q} \approx - 2 {\bf Q} = {\bf Q}$ .\newline
Expanding the function $\gamma({\bf q})$ around these 3 points, 
and assuming the order of limits $w \gg g \gg w q^2$, we find
\be\label{approx_int}
\begin{split} &
{\mathcal H}_{\textrm{eff}} \approx \\ &
- \frac{U_{b f}^2}{N^2} N_0 \sum_{\bf q \approx 0} 
\Big[
\left( \frac{1}{4 g} + \frac{1}{12 w} \right)
\left(
F_1^{\dagger} F_1 + F_2^{\dagger} F_2
\right)
\\ &
+
\left( \frac{1}{4 g} - \frac{1}{12 w} \right)
\left(
F_1^{\dagger} F_2 + F_2^{\dagger} F_1
\right)
\Big] 
\\ &
- \frac{U_{b f}^2}{N^2} N_0 \frac{1}{3 w}  \sum_{\bf p \approx 0} 
\left( F_1^{\dagger} F_1 + F_2^{\dagger} F_2 
\right)
|_{{\bf q} = \pm {\bf Q} + {\bf p}}
\end{split}
\ee

The case of non-interacting bosons $g=0$, corresponds to a different limit than above 
$w \gg w q^2 \gg g=0$. The Bogoliubov spectrum becomes the same as the band structure $E \rightarrow \epsilon$,
the Bogoliubov transformation parameters simplify to $u=1, v=0$,
and the long-wavelength limit gives
\be
V_{1 1}({\bf q}) = V_{1 2}({\bf q}) =  U_{b f}^2 \frac{2 \rho_0}{3 w} \frac{1}{q^2}
\; .
\ee
We find that in this limit the effective interactions are \emph{long-range} in real space,
and attractive.

The boson interaction strength $g$ controls the range for the effective interaction $\ell \sim 1/g$.
Despite us using a weak interaction limit in the Bogoliubov theory, when assuming $w \gg g$ we end 
up having the boson interaction coupling $g$ dominating the nature of the effective interaction, 
with attractive ${\bf q} \approx 0$ interactions.
The diagonal terms of the interaction vertex correspond to real space interactions between 
sites on the same sublattice. This type of interaction includes on-site interactions, which 
are expected to be the strongest of this kind.
In contrast, the off-diagonal (the $F_1^{\dagger} F_2$ term) terms of the interaction vertex 
correspond to real space interactions between sites on \emph{different} sublattices. The shortest 
range interactions in this class are nearest neighbor interactions. Since the off-diagonal and 
diagonal terms are comparable in magnitude, we expect an attractive effective nearest neighbor 
of comparable strength to that of the effective on-site interaction.

From the features of the effective interaction vertex, we are lead to believe that 
the phenomenological model of Ref.~\onlinecite{Uchoa:prl07} may be an appropriate
description of this system, which considers fermions on the
honeycomb lattice, with only on-site and nearest neighbor interactions,
parametrized by $g_{1,2}$ respectively. For the range
of parameters we find in the present work $g_2 < 0$, and 
depending on the strength of $U_{ff}$, the bare on site repulsion, we can 
have either positive or negative sign of $g_1$. Specifically, we can realize $g_1 <0$ when 
$U_{ff} \ll \frac{U_{b f}^2 N_0}{N^2} \left( \frac{1}{4 g} + \frac{1}{12 w} \right)$,
and $g_1 >0$ when $U_{ff} \gg \frac{U_{b f}^2 N_0}{N^2} \left( \frac{1}{4 g} + \frac{1}{12 w} \right)$.

Uchoa \etal \cite{Uchoa:prl07} find via a mean field analysis that the ground state
may be a $p+ip$ superconducting state for $g_2<0$ and $g_1<0$
and a mixed s-wave and $p+ip$ superconducting state for $g_2 <0$ and $g_1>0$.
Therefore it would seem that for strong bare fermion repulsion $U_{ff}$, one should expect 
the $p_x +i p_y$ phase.


\section{Magnetic field splitting}
\label{Magnetic_Field_Splitting}

Within the continuum Dirac theory we found zero modes in all the geometries we considered, 
topological protection occurs modulo symmetry mandated degeneracy (see section~\ref{vortex_edge_equiv}). 
The 4-fold degenerate zero modes we found in the two spin-singlet phases
are protected as a result of the 4-fold degeneracy mandated by the $SU(2)$
symmetries of the spin and the Dirac valley spinor.
As first pointed out in ref.~\onlinecite{Ghaemi:2007}, a Zeeman field is found to
split the spin-degenerate zero modes of the s-wave phase into Zeeman pairs, 
with a splitting proportional to magnetic field, at first order in perturbation theory.
The magnetic field explicitly breaks the $SU(2)$ spin symmetry, and therefore the zero modes can now split.
In our approach, it is easy to show this is an \emph{exact} result, and that the zero mode states remain 
exact eigenstates of the system, albeit with a nonzero energy.

Without a magnetic field the system is isotropic in the spin sector, and so we are free to choose
the direction of the magnetic field in spin space. It is convenient to choose the Zeeman field in the y-direction. 
The zero modes we found satisfy $\sigma^y \psi = \sigma \psi$, and so we have precisely
$B \sigma^y \psi = B \sigma \psi$ splitting the zero modes.
It is important to point out though, that while the mathematical spectrum of the BdG Hamiltonian
has 4 zero modes splitting in to 2 positive and 2 negative, the physical excitations of the system 
consists only of the positive energy states (because of $\omega^x B \sigma^y \omega^x = B \sigma^y$, 
the BCS Hamiltonian identity $\omega^x {\mathcal H} \omega^x = -{\mathcal H}^*$ still holds),
and so there will be a doublet of lowest energy excitations, with $E = B$.

Unlike a charged fermion superconductor, in a fermionic superfluid there is no need for 
a magnetic field to create vortices. In any experimental cold atom apparatus, any magnetic 
field can be made extremely small. Therefore, it would be a great advantage to realize the
spin-singlet phases we discuss in this manuscript in a cold atom system, 
rather than a solid state system. If the magnetic field is too weak to destroy the fermion pairing, 
the only important effect of the magnetic field is to introduce a Zeeman field coupling to the 
spin degrees of freedom, and the spectrum of vortex-core bound states can be manipulated.

\begin{figure}
	\centering
		\includegraphics[width=2.0in]{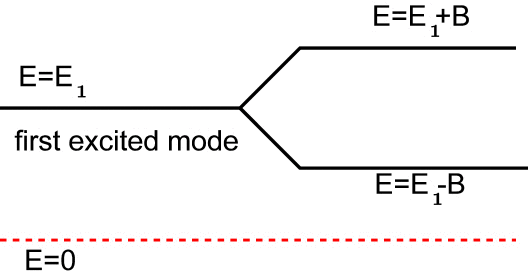}
	\caption{Influence of the Zeeman field splitting on the excitation spectrum 
	in the case that vortex core zero modes are absent}
	\label{fig:RF1}
\end{figure}

\begin{figure}
	\centering
		\includegraphics[width=2.0in]{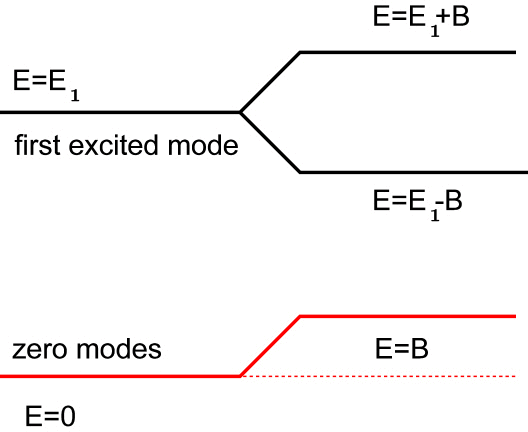}
	\caption{Influence of the Zeeman field splitting on the excitation spectrum
	in the case that vortex core zero modes exist}
	\label{fig:RF2}
\end{figure}

Using the Zeeman field splitting of the bound states, we now proceed to propose an experiment to 
probe whether zero modes exist in these systems (as found in the continuum Dirac theory)
or not (expecting the zero modes to split slightly as found in the numerics of Section~\ref{Numerics} ).
The magnetic field allows us to control the low energy 
spectrum of the system, and this will possibly make it easier to identify in RF (low frequency) 
absorption measurements. If we assume all energy states of the experimental system are Kramers doublets, 
in the absence of zero modes, the lowest energy excitation ($E_1$) will be lowered
when applying a magnetic field $E = E_1 - B$ (see Fig.~\ref{fig:RF1}). When zero modes exist, then when applying a magnetic field the lowest
excitation energy will move up in energy ($E_0 = B$) (see Fig.~\ref{fig:RF2}). The lowest excitation energy in the system will then \emph{decrease} with rising magnetic field in the absence of near-zero modes, but will \emph{increase} 
if they exist in the system.
This serves an experimental method to identify the existence of these states, which could easily be carried out in 
cold atom systems, but are perhaps more difficult in superconducting solid states systems.

\section{Discussion and Conclusion}
\label{discussion}


In this manuscript we have explored whether topological zero modes
exist in a number of possible fermionic condensate phases on the honeycomb lattice.
We {\bf examined 2 spin-singlet phases both of} which are fully gapped in the entire Brillouin zone.
We have found that 4-fold degenerate topological zero modes exist within the continuum Dirac theory,
{\bf for these two phases}.
We have done this by explicitly solving for zero modes bound to vortex cores, at sample edges,
and in SNS junction geometries. In all cases, the edge state and vortex core calculations agree completely,
with the same degeneracy of zero modes being found, but in the SNS junction geometry an extra accidental 
symmetry doubles the number of zero modes from 4 to 8. 

With an even degeneracy of vortex core zero modes, the majorana zero modes are not compelled to
pair into fermionic degrees of freedom between spatially remote vortices, but rather \emph{locally}
at each vortex core. The natural mechanism for entanglement of vortex pairs is therefore lost,
and no non-Abelian statistics between vortices should appear.

The topological zero modes existence crucially depends on the 
emergent low energy $SU(2)$ symmetry of the Dirac valley spinor structure in the Dirac theory.
This symmetry does not strictly hold in the original lattice model, and this brings up the 
possibility that the seemingly protected zero modes found in the Dirac theory,
are split in the full honeycomb lattice model. 
While corrections to the Dirac theory give an unclear picture of the fate of the zero modes when the effective low energy symmetry is broken, the numerical diagonalization
we have performed on the lattice model confirm that indeed this is the case - the zero modes are
split. In the present context simply using the continuum Dirac model is inaccurate - even when
the vortex structure is slowly varying on the scale of the lattice, the Dirac theory is still only approximate,
and in fact the zero modes are \emph{not} topologically protected.

We have also discussed the Zeeman field splitting of the vortex core bound states in this phases.
We suggested an experiment taking advantage of this splitting to ascertain whether zero modes 
exist or not in this system, by tracking how the excitation energies change when modifying the Zeeman field. 

In order to realize the experiment we propose, one first needs to create a condensate on the honeycomb 
lattice. It may be possible to realize a superconducting state in graphene by superconducting leads 
inducing electron pairing via the proximity effect. Another possibility we have discussed, is forming fermion condensates in cold atom Bose-Fermi mixtures. The latter however will most probably require the fermions 
to be cooled down to very low temperatures compared with the energy scales of optical lattices,
which is challenging in current experiments. 
A cold atom gas is perhaps the ideal realization of the condensate for our proposed experiment,
since magnetic fields are not involved in the forming of vortices in the first place, and so can be freely
manipulated, without affecting the condensate or the vortices too much.

In the context of the zero modes in the ``ordinary'' $p_x + i p_y$ state, 
it has already been suggested to probe the bound state spectrum by RF absorption\cite{Stern:prb07},
and STM measurements\cite{Bolech:PRL07}. The same tools could be used to probe the bound state
spectrum in the phases we discuss here.
As opposed to Refs.~\onlinecite{Stern:prb07,Bolech:PRL07}, in the experiment we propose one would be looking 
for how the spectrum moves when changing the magnetic field, rather than simply looking at a static spectrum.
It is sometimes more easy to notice something that is moving, 
rather than stationary, and so it may prove easier to detect the spectrum shifts.

The reaction to magnetic field of the bound states is not limited to vortex cores - it could be 
discernible in edge states as well if the sample is small enough that the discreteness of the 
energy levels bound to the edge becomes evident. The edge state spectrum in all the phases we
consider here is always linear 
in the transverse momentum $E \sim q v$, with some effective velocity $v$. For a finite system,
the momentum will be quantized, $q = \frac{\hbar}{2 \pi L} (\ell + \gamma)$, and the bound states have 
a level spacing of $\sim \frac{\hbar v}{2 \pi L}$. 
If the Zeeman splitting as well as the thermal energy scale
are smaller than this level spacing, the effect we describe here is in principle observable.
In practice, the experimental probe must be sensitive enough to probe these small energy scales.

Finally, our present work has forced us to generalize some ideas that were understood and developed
in the context of the ``ordinary'' $p_x + i p_y$ state.
Topological zero modes were previously understood to be topologically protected only 
if \emph{single} zero modes existed (in a unit vorticity vortex). We have generalized this view of topological 
protection to accommodate symmetry mandated degeneracy of quasiparticle excitations,
for which the phases of the continuum Dirac theory we considered are examples .
Our conclusion is that zero modes can be protected to all perturbations that preserve 
the symmetries, and as such are protected by the combination of symmetry and topology. 
If the symmetries are explicitly broken by a perturbation, then the zero modes may split. 
This is precisely the reason the Dirac theory and the precise honeycomb lattice model differ,
and also the reason for the Zeeman splitting of the zero modes in the Dirac theory.
This synergistic protection, while clearly 
more fragile than the topological protection of a single zero mode may prove important in 
understanding many other physical systems beyond those we discuss here.

We have also generalized the known connection between topological zero modes bound
to vortex cores and at sample edges in the well studied ``ordinary'' $p_x +i p_y$ state. 
We have shown that in quite general settings, with the possibility of symmetry mandated degeneracy
included, zero modes bound to vortices and edges should be \emph{identified}.

Our analysis was limited to a number of presumed pairing states for fermions
on the honeycomb lattice, but similar phenomena may be uncovered in other states 
involving the honeycomb lattice. In particular Refs.~\onlinecite{Herbut:prl07,Chamon:prl07} 
have discussed non-condensate models on the honeycomb lattice with vortices possessing
zero mode bound states in their core. In both cases only a vortex calculation was carried out, 
and given our general observation that vortex core bound zero modes should be identified with 
edge state zero modes, we expect these zero modes to appear at sample edges in the models of 
Refs.~\onlinecite{Herbut:prl07,Chamon:prl07}. 
In Ref.~\onlinecite{Chamon:prl07} the authors find there is a \emph{single} zero mode 
bound to the vortex core.
Another related model which exhibits topological zero modes at sample edges is the 
Kane-Mele model\cite{Kane:prl05},
which both in the precise lattice model, as well as in a continuum limit\cite{Sengupta:prb06}, exhibits edge state zero modes (in the continuum case, for an armchair boundary - the zigzag
boundary suffers from the same problems we pointed out in section~\ref{Edge_states}).
Finally, we mention a very recent publication\cite{seradjeh-2008} finding zero modes bound to vortices in a 
bilayer-graphene exciton condensate. As in our case, the zero modes turn out to split in the precise lattice model.
We suspect that other interesting possible states of matter on the honeycomb lattice geometry exist, as well 
as in 3-dimensional geometries that supply the common ingredient in all these models - the Dirac nodes
in the lattice band structure. The physics of a 3-dimensional version of the Kane-Mele 
model\cite{Fu:prl07} is realized in Bi$_{1-x}$Sb$_x$, as recently probed in Ref.~\onlinecite{Cava:nature08},
and following this work Ref.~\onlinecite{fu:096407} has suggested that majorana fermion zero modes should appear 
at the interface between a topological insulator and an s-wave superconductor.

\begin{acknowledgments} 
 
We would like to acknowledge C. Chamon, D. Novikov, N. Read,
R. Sensarma, for illuminating discussions. This work was supported by the NSF through grant DMR-0803200 
and through the Yale Center for Quantum Information Physics.

\end{acknowledgments}

\appendix

\section{Spinless p+ip condensate}
\label{app:spinless}

In this appendix we analyze the spinless $p_x + i p_y$ phase mentioned briefly in the main text.
We will find that this phase in some geometries will posses zero modes, but these are \emph{bulk states}
rather than bound states.

\subsection{SNS junction}
\label{spinless_SNS}

In this subsection we will analyze the SNS junctions in the spinless $p_x + i p_y$ phase.
We consider here only wavefunctions that are uniform in the direction parallel to the SNS
junction boundaries, and find in stark contrast to the spin-singlet phases, that no zero modes
exist.
The steps of our analysis follow closely those of the SNS calculations for the spin-singlet phases, 
and so we will describe our calculations in minimal detail.

With a (piecewise) uniform pairing function, combining the kinetic energy 
\eqref{H_0} and the spinless $p_x+i p_y$ pairing term \eqref{H_3}
yields a BdG equation of the form
\be\label{spinless_BdG}
\begin{split} &
{\mathcal H}_{BdG} \psi = E \psi \\ = &
\Big[ 
\mu \omega^z - i v {\hat D}
+
\tau^x \Delta \eta^z
{\hat D}
\left[ \omega^x \cos(\phi) + \omega^y \sin(\phi) \right] \eta^y \Big] \psi
\; ,
\end{split}
\ee
where we have dropped the $2$ subscript from both the order parameter phase and magnitude, to avoid clutter.

We consider only states that are uniform in the direction parallel to the SNS junction boundaries,
and use the same unitary transformation $U(\alpha)$ to rotate the angle between the SNS
boundaries and the y-axis $\alpha \rightarrow 0$, in the operator ${\hat D}(\alpha)$
appearing in both the kinetic energy and the pairing term. All other parts of the BdG 
Hamiltonian remain invariant, and as long as we ignore the quadratic correction \eqref{H_4},
we can simply set $\alpha = 0$. The BdG equations reduce to 
\be
\Big[ 
\mu \omega^z - E - i v \eta^x \partial_x
+ \tau^x \Delta
i \partial_x
\left[ \omega^x \cos(\phi) + \omega^y \sin(\phi) \right] \Big] \psi = 0
\; .
\ee
As before, we will work in the London gauge which we can get by applying the unitary 
transformation ${\mathcal O} = e^{-i \frac{\phi}{2} \omega^z} $. We are left with
\be
\Big[ 
\mu \omega^z - E - i v \eta^x \partial_x
+ \tau^x \Delta
i \partial_x \omega^x \Big] \psi = 0
\; ,
\ee
from which it is clear that we can choose solutions that are eigenstates of both $\eta^x$ and $\tau^x$,
such that $\eta^x \psi = \eta \psi$ and $\tau^x \psi = \tau \psi$. The BdG equations then can be 
reorganized in the form $i \partial_x \psi = A \psi$ with
\be
A = 
\frac{1}{v^2 - \Delta^2}
\left(
\begin{array}{ll}
 v \eta  (E -\mu ) & \Delta  (E +\mu ) \tau  \\
 \Delta  (E -\mu ) \tau  & v \eta  (E +\mu )
\end{array}
\right)
\; .
\ee
The eigenvalues of the matrix $A$ are 
$v E \eta \pm \sqrt{\Delta^2 E^2 + \mu^2 (v^2 - \Delta ^2) }$,
and since we expect $v \gg \Delta$, the eigenvalues will in general be \emph{real}
numbers. As a consequence, we can only have
solutions of $\psi$ that are exponentials of 
purely imaginary arguments. As a result, no 
bound states can appear - these require some exponential 
decay of the wavefunction. 

\subsection{Edge geometry}

We will now address the edge geometry in the spinless $p_x + i p_y $ phase.
Starting form the BdG equation \eqref{spinless_BdG}, with the phase chosen as $\phi =0$,
and assuming a transverse momentum $q$, such that $\psi = e^{i q y} \psi(x)$.
The reduced BdG equations are
\be
\left[ 
\mu \omega^z - i v {\hat D}
+
\tau^x \Delta \eta^z {\hat D} \omega^x \eta^y - E \right] \psi = 0
\; ,
\ee
with ${\hat D} = \eta^x \partial_x + \omega^z \eta^y \partial_y$.
Reorganizing the BdG equations yields
\be
\left[ 
\mu \omega^z - E + v q \eta^y \omega^z - q \Delta \tau^x \eta^z \omega^y \right] \psi
=
i \left[ v \eta^x - \Delta \tau^x \omega^x \right] \partial_x \psi
\; .
\ee
Using the identity 
$\left[ v \eta^x - \Delta \tau^x \omega^x \right]^{-1} = 
\frac{1}{v^2 - \Delta^2}
\left[ v \eta^x + \Delta \tau^x \omega^x \right]$, we bring the equation to the form 
$\partial_x \psi = A \psi$, with
\be
A = -i \left[
\frac{1}{v^2 - \Delta^2} \left(
v \eta^x + \Delta \tau^x \omega^x 
\right) \left( \mu \omega^z - E \right)
+ i q \omega^z \eta^z
\right]
\; .
\ee

We are free at this stage to choose solutions that are eigenstates of $\tau^x \psi = \tau \psi$,
so we simply replace $\tau^x \rightarrow \tau$. Furthermore, it is useful at this point to 
denote $z = \frac{x \mu}{\sqrt{v^2 - \Delta^2}}$, $\epsilon = \frac{E}{\mu}$ and 
$k = \frac{q \sqrt{v^2 - \Delta^2}}{\mu}$, all of which are dimensionless quantities.
We can also assume without loss of generality that $\mu>0, v>\Delta$, and $E \geq 0$.
The BdG equations now become $\partial_z \psi = {\tilde A} $ with
\be
{\tilde A} = -i \left[
\frac{1}{\sqrt{v^2 - \Delta^2}} \left(
v \eta^x + \Delta \tau^x \omega^x 
\right) \left( \omega^z - \epsilon \right)
+ i k \omega^z \eta^z
\right]
\; .
\ee
The 4 eigenvalues of the matrix $i {\tilde A}$ in this notation are
\be
\begin{split} &
\pm \lambda_{\eta} = 
\\ &
\pm \left[
\left( 1 - k^2 \right)
+ \epsilon^2 \frac{v^2+\Delta ^2}{v^2-\Delta ^2}
+ \eta 
2 v \epsilon 
\frac{\sqrt{ \left( v^2 - \Delta^2 \right) + \Delta ^2 \epsilon ^2 }}{v^2-\Delta ^2}
\right]^{1/2}
\; ,
\end{split}
\ee
where $\eta = \pm 1$.

We are interested in exploring zero modes, so at this point we set $\epsilon = 0$,
to identify which eigenvalues can give solutions that are exponentially decaying in the $z>0$
region. The eigenvalues of ${\tilde A}$ become $\pm \lambda_{\eta} = \mp i \sqrt{ 1 - k^2 } $. 
For exponentially decaying zero modes, we must therefore have $|k| >1$.
In this regime, $q > \frac{\mu}{\sqrt{v^2 - \Delta^2}}$.
It is noteworthy that the nodes in the bulk spectrum for this phase occur
in the continuum theory at precisely ${\bf q} = \left( \sqrt{3} , 1 \right) \frac{\mu}{\sqrt{v^2 - \Delta^2}}$.
Therefore, these bound states may be identified with the bulk zero modes.

\subsection{Vortex geometry}

In this subsection we turn to explore whether zero modes exist bound to vortex cores in the 
spinless $p_x + i p_y$ phase. 

We start with the BdG equation $\left[ {\mathcal H}_0 + {\mathcal H}_3 \right] \psi = E \psi$
with ${\mathcal H}_{0,3}$ from \eqref{H_0}, and \eqref{H_3}. We choose the eigenstates to 
satisfy $\tau^x \psi = \tau \psi$, as in the SNS and edge geometries.
We will model the vortex by assuming the form $i \Delta_2 = \Delta(r) e^{+i \phi}$, with $\Delta(r)$ real
(as for the spin singlet $p_x+ip_y$ case, the convention is different from earlier sections so that 
we can use the same ansatz for the polar angle dependence as for the s-wave case). 
Also, since it will prove convenient, we will assume that the order parameter radial profile is piecewise 
uniform. 

As in previous subsections,
we find the $\phi$ dependence can be eliminated from the zero-mode problem by 
choosing the wavefunction form
\be
\psi(r,\phi) = e^{i \ell \phi} \left(
u_1(r), e^{i \phi } u_2(r),v_1(r),e^{-i \phi } v_2(r)
\right)^T
\; .
\ee
The reduced ODEs then involve only the radial coordinate, and can be cast
in the form $\partial_r \psi = A \psi$ where
\be
\begin{split} &
A =
\frac{1}{r} \left( \ell \omega^z \eta^z - \frac{1}{2} \right)
\\ &
+ \frac{1}{v^2 - \Delta^2}
\Big[
\frac{v^2}{2 r} \eta^z
+ i v \eta^x (E - \mu \omega^z)
\\ &
- \frac{v \Delta \tau}{2 r} i \omega^x \eta^y
+ \Delta \tau (E i \omega^x - \mu \omega^y)
\Big]
\; .
\end{split}
\ee
For the purpose of showing that no bound state zero modes exist, it will suffice to consider the asymptotic limit 
($r \rightarrow \infty$) alone.
We neglect all terms in $A$ that have a factor $1/r$
\be
A =
\frac{1}{v^2 - \Delta^2} \left[
i v \eta^x (E - \mu \omega^z)
+ \Delta \tau (E i \omega^x - \mu \omega^y)
\right]
\; .
\ee
Setting $E = 0$, we find $A$ has the eigenvalues 
$\pm i \frac{\mu}{\sqrt{v^2 - \Delta^2}}$, which are purely imaginary.
Therefore no bound states with zero energy are allowed! Thus, we conclude that no zero modes exist
at the vortex core in this phase.

This particular calculation shows us that having a Dirac equation structure in the BdG equations, 
is \emph{not} a sufficient condition for topological zero modes to be present.

\subsection{Numerics}
\label{app:numerics}
 
In the previous sections of this appendix, we found that the edge state geometry can
support some zero modes with a wavefunction concentrated at the at the edge.
In contrast, in the vortex calculation found no bound states zero modes.
As we argued in this paper, in a fully gapped system, we expect a general correspondence 
between the edge state spectrum and the vortex core bound state spectrum. This expectation 
does not hold here, presumably due to the fact that this phase is \emph{not} fully gapped.

To verify that the vortex calculation result is correct (it uses the approximate continuum
description) we employed the numerical methods of section~\ref{Numerics}.
Using the precise lattice pairing function for the spinless $p_x+ip_y$ phase, 
with a square lattice patch of $1824$ lattice sites, we calculated 
the low energy spectrum for the vortex state, with two different vortex core sizes $R=0,\sqrt{5}$.
We set the parameters $|\Delta| = 0.5$ and $\mu = 0.4$.
The lowest energies divided by the de Gennes scale are $\frac{E}{\frac{\Delta^2}{E_F}} = 
\pm 0.0270939, \pm 0.0558193, \pm 0.0765734 \ldots$, for $R=0$ and
$\frac{E}{\frac{\Delta^2}{E_F}} =
\pm 0.0234966,  \pm 0.0350545, \pm 0.0630661 \ldots$ for $R=\sqrt{5}$.
The results in both cases are similar - we find low energy states exist, far below the 
de Gennes energy scale, but \emph{all} the states with energy below the de Gennes scale, are 
delocalized bulk states and \emph{not} concentrated near the vortex core. This result would 
indicate that the vortex calculation and edge state calculations are not at odds - the zero 
modes found in the edge state calculation are related to the bulk low energy states
that appear due to the nodes in the pairing function.


\end{document}